%% file: BSM+Composites.tex
\documentclass[aps,prd,reprint,superscriptaddress,floatfix,nofootinbib]{revtex4-1}

\usepackage{microtype}
\usepackage{comment}

\usepackage{graphicx}
\usepackage[dvipsnames]{xcolor}
\usepackage{rotating}
\usepackage[caption=false,singlelinecheck=false]{subfig}
\usepackage{amsmath}
\usepackage{amssymb}
\usepackage{multirow}
\usepackage{amsfonts,dsfont}
\usepackage{bm}
\usepackage{physics}
\usepackage{mathtools}
\usepackage{slashed}

\usepackage{booktabs}
\usepackage{array}
\usepackage{enumitem}
\usepackage[normalem]{ulem}

\usepackage{url}
\usepackage{xspace}
\usepackage{siunitx}
\usepackage{xfrac}
\usepackage{hyperref}
\usepackage[nameinlink]{cleveref}
\usepackage{appendix}

\usepackage{xifthen}
\usepackage{xcolor}
\hypersetup{colorlinks,	linkcolor={blue!75!black},	citecolor={blue!75!black},	urlcolor={blue!75!black}}

\newcommand{\imag}{\text{i}}

\def\s0#1#2{\mbox{\small{$ \frac{#1}{#2} $}}}
\def\0#1#2{\frac{#1}{#2}}

\newcolumntype{x}[1]{>{\centering\arraybackslash\hspace{0.cm}}p{#1}}

\DeclareSymbolFont{symbolsC}{U}{pxsyc}{m}{n}
\newcommand{\commentmute}[1]{} 

\usepackage{tikz}
\usepackage{bm}
\usetikzlibrary{decorations.pathmorphing}

\tikzstyle{dashed}=                  [dash pattern=on 4pt off 1.5pt]

\tikzset{cross/.style={cross out, draw=black, minimum size=2*(#1-\pgflinewidth), inner sep=0pt, outer sep=0pt},
	cross/.default={1pt}}
\usetikzlibrary{decorations.pathmorphing,decorations.markings}
\pgfdeclaredecoration{complete sines}{initial}
{
	\state{initial}[
	width=+0pt,
	next state=sine,
	persistent precomputation={\pgfmathsetmacro\matchinglength{
			\pgfdecoratedinputsegmentlength / int(\pgfdecoratedinputsegmentlength/\pgfdecorationsegmentlength)}
		\setlength{\pgfdecorationsegmentlength}{\matchinglength pt}
	}] {}
	\state{sine}[width=\pgfdecorationsegmentlength]{
		\pgfpathsine{\pgfpoint{0.25\pgfdecorationsegmentlength}{0.5\pgfdecorationsegmentamplitude}}
		\pgfpathcosine{\pgfpoint{0.25\pgfdecorationsegmentlength}{-0.5\pgfdecorationsegmentamplitude}}
		\pgfpathsine{\pgfpoint{0.25\pgfdecorationsegmentlength}{-0.5\pgfdecorationsegmentamplitude}}
		\pgfpathcosine{\pgfpoint{0.25\pgfdecorationsegmentlength}{0.5\pgfdecorationsegmentamplitude}}
	}
	\state{final}{}
}

\tikzset{
	vertex/.style={draw,shape=circle,fill=black,minimum size=5pt,inner sep=0pt},
	scalar/.style={ thick, dashed,  draw=black,decorate},
	scalarthick/.style={  draw=black,decorate, line width=2pt},
	gauge/.style={ thick, draw=black,decorate, decoration={snake, amplitude=1.25pt, segment length=2.5pt}},
    gaugeSM/.style={ thick, draw=blue,decorate, decoration={snake, amplitude=1.5pt, segment length=4.pt}},
	gluon/.style={decorate, thick, draw=black, decoration={coil,amplitude=2.7pt, segment length=2.75pt}},
	fermion/.style={line width= 0.7 pt,  draw=black, postaction={decorate}, decoration={markings, mark=at position .575 with {\arrow[draw=black,line width=1 pt]{>}}} },
	fermionSM/.style={draw=blue, line width= 1.2 pt, postaction={decorate}, decoration={markings, mark=at position .575 with {\arrow[draw=blue]{>}}} },
	fermionplain/.style={ thick, draw=blue },
	comp/.style={ draw=black, line width= 1. pt, double, dashed,
  postaction={decorate,draw=black},
  decoration={markings, mark=at position .65 with {\arrow[draw=black,line width=2. pt]{>}}}
  } 
}
\graphicspath{{../figures/},{figures/}}
\begin{document}
	
	%%%%%%%%%%%%%%%%%%%%%%%%%%%%%%%%%%%%%%%%%%%
	% opening
	%%%%%%%%%%%%%%%%%%%%%%%%%%%%%%%%%%%%%%%%%%%
	\title{Flavour hierarchies from emergent fundamental partial compositeness}

    \author{Florian~Goertz}
	\affiliation{Max-Planck-Institut f\"ur Kernphysik P.O. Box 103980, D 69029, Heidelberg, Germany}
	\author{\'Alvaro~Pastor-Guti\'errez}
	\affiliation{Max-Planck-Institut f\"ur Kernphysik P.O. Box 103980, D 69029, Heidelberg, Germany}
	\author{Jan~M.~Pawlowski}
	\affiliation{Institut für Theoretische Physik, Universität Heidelberg, Philosophenweg 16, 69120 Heidelberg, Germany}
	\affiliation{ExtreMe Matter Institute EMMI, GSI Helmholtzzentrum für Schwerionenforschung mbH, Planckstr.\ 1, 64291 Darmstadt, Germany}

%%%%%%%%%%%%%%%%%%%%%%%%%%%%%%%%%%%%%%%%%%%%%
\begin{abstract}
Composite Higgs extensions of the Standard Model provide an explanation for the large hierarchies between the Yukawa couplings. We study their realisation in the context of fundamental partial compositeness where the Standard Model fermions mix linearly with bound states of the new sector, consisting of a fermion and a scalar. The properties of this composite are unravelled with the functional renormalisation group approach using dynamically emergent composites. Specifically, we extract the scaling of correlation functions and provide indicative estimates for the minimal incarnation of the theory. 
\end{abstract}
%%%%%%%%%%%%%%%%%%%%%%%%%%%%%%%%%%%%%

\maketitle

%%%%%%%%%%%%%%%%%%%%%%%%%%%%%%%%%%%%%%%
\section{Introduction}
\label{sec:Introduction}

Understanding the hierarchical structure of fermion masses and mixings is one of the major open questions in fundamental physics. 
The concept of partial compositeness (PC) for fermions, first proposed in \cite{Kaplan:1991dc} and put forward in \cite{Contino:2004vy,Agashe:2004rs,Contino:2006qr,Contino:2003ve} in extra-dimensional duals of composite Higgs (CH)~\cite{Kaplan:1983fs,Kaplan:1983sm,Dugan:1984hq} models, offers a promising means to address this question. In this approach, the fermion mass terms of the Standard Model (SM) are generated from linear mixings of SM-like fermions of each flavour with composite fermionic operators ${\cal O}_B$, containing fundamental fields of a new sector that are bound together by a novel confining interaction. Below the condensation scale $\Lambda_c$, these terms lead to the light fermion mass eigenstates being a superposition of elementary SM-like fermions and composite resonances excited by the operators ${\cal O}_B$ in the infrared (which explains the denotation `partial compositeness'). The latter resonances provide the connection to the composite Higgs and thus to electroweak symmetry breaking (EWSB). 

Small differences in the scaling dimensions of the composite operators translate to exponentially large differences in the strengths of the linear mixings at low energies due to the renormalisation-group evolution from a large UV flavour scale $\Lambda_{\rm UV}$, where the couplings of the SM fermions with the strongly coupled sector are generated, down to the condensation scale $\Lambda_c$. Assuming an almost conformal behaviour between those scales, the linear-mixing coefficient will scale as $\sim (\Lambda_c/\Lambda_{\rm UV})^{d-5/2}$, with the dimension $d=[{\cal O}_B]$ of the composite operator, see e.g.\cite{Contino:2010rs}. After integrating out the heavy states, these hierarchically different couplings to the bound states of the new strong sector lead to hierarchical mass eigenvalues of quarks and leptons. 

While the initial focus in the literature was on effective low-energy descriptions of PC (see \cite{Contino:2010rs,Bellazzini:2014yua,Panico:2015jxa} for reviews), more recently UV-complete realisations have been explored \cite{Barnard:2013zea,Ferretti:2013kya,Ferretti:2014qta,Cacciapaglia:2014uja,Vecchi:2015fma,Sannino:2016sfx,Cacciapaglia:2017cdi,Agugliaro:2019wtf,Cacciapaglia:2020kgq}. These consider the fundamental degrees of freedom and the dynamics that lead to the bound states that mix linearly with the elementary SM-like fermions. Here, an obvious approach is to assume the composite fermions being composites of three fundamental fermionic degrees of freedom \cite{Barnard:2013zea,Ferretti:2013kya,Ferretti:2014qta,Vecchi:2015fma}. Such constructions however face severe challenges since the scaling dimension of the composite operators needs to deviate very significantly from the canonical value of $[{\cal O}_B]_{0,\,c} =3 [{\cal F}]_c=9/2$. The latter would lead to fermion masses, that are too small. The linear mixings are suppressed by $[{\cal O}_B]_{0,\,c}+3/2 - 4 = 2$ powers of the flavour scale $\Lambda_{\rm UV}$ where they emerge as higher dimensional operators involving fundamental fermions (flavour bounds indicate $\Lambda_{\rm UV} \gg \Lambda_c$). This issue is particularly severe in the case of the large top-quark mass.

In fact, lattice results suggest that large deviations from the canonical scaling dimension (i.e., large anomalous dimensions) are not possible for the straightforward realisation with three-fermion bound states~\cite{DeGrand:2015yna, Pica:2016rmv, Ayyar:2018glg, BuarqueFranzosi:2019eee}. 
Thus, it is not obvious whether this completion improves the situation compared to the technicolour (TC) way of generating fermion masses via bilinear couplings to the EWSB condensate - where in general the masses are also suppressed too much (unless the scaling dimension of the Higgs-like bound state becomes as small as to re-introduce UV instabilities), see e.g.~\cite{Contino:2010rs}. In consequence, finding alternative mechanisms of realising PC becomes a priority. 

One of these alternatives is based on the assumption that the fermionic composites are not formed by three elementary fermions but by an elementary fermion ${\cal F}$ and a scalar ${\cal S}$, see~\cite{Sannino:2016sfx,Cacciapaglia:2017cdi,Agugliaro:2019wtf}. These scenarios have been dubbed ``fundamental partial compositeness" (FPC), and rely on renormalisable Yukawa-like couplings of the SM-like fermions to ${\cal S}$ and ${\cal F}$. The latter two form a bound state ${\cal B} \sim {\cal S F}$ that mixes with the SM fermions in the infrared (IR). Here, ${\cal B}$ refers to the lightest resonance in a tower of states excited by ${\cal O}_B$ \cite{Cacciapaglia:2020kgq}. 

The inclusion of scalars and hence the resulting canonical dimension of $[{\cal O}_B]_{0,\,c} =[{\cal F}]_c+[{\cal S}]_c=5/2$ carry the crucial advantage that the linear mixings are not necessarily suppressed. Thus, the hope is to obtain the heavy top quark mass, while the light fermion masses are still set via natural input parameters. The presence of elementary scalars may reintroduce a hierarchy problem, but the approach could be seen as an intermediate step to a full UV theory. In fact, an interesting construction circumventing naturalness problems could be two-step models of condensation where the scalars ${\cal S}$ would emerge from fundamental fermions in the UV.

In the original works on FPC, the question of a dynamical generation of the (hierarchically) light fermion masses was left open~\cite{Sannino:2016sfx,Cacciapaglia:2017cdi,Agugliaro:2019wtf,GiacomoPrivComm}. In principle, one can just generate such hierarchical fundamental Yukawa couplings around the condensation scale or from hierarchies in the fundamental scalar masses \cite{Cacciapaglia:2020kgq}, however, this would not address the flavour puzzle.
Given the promising starting point discussed above, it is worthwhile to investigate whether a natural dynamical emergence of all fermion masses could be realized using FPC. 
To this end, in this work we explore the possible range of anomalous dimensions of the ${\cal O}_B$ operators, containing a scalar and a fermion, to see if an envisaged evolution over a sizeable almost-conformal regime could generate the sought flavour hierarchies. 

Studying the dynamics of strongly coupled theories and the emergent formation of composites is a non-trivial task requiring non-perturbative methods. Usually, this is addressed with Monte-Carlo simulations on the Lattice. In the present work we use the functional Renormalisation Group (fRG)~\cite{Wetterich:1992yh,Ellwanger:1993mw, Morris:1993qb} to study the anomalous scaling of the linear mixing coupling between SM fermions and composites. Functional methods based on the effective action allow for a systematic treatment of non-perturbative effects while being versatile enough to scan over the parameter and model sets. This flexibility suits the needs for investigations of strong and non-perturbative new physics. Specifically, the fRG allows for the systematic inclusion of emergent composites, see \cite{Gies:2001nw, Pawlowski:2005xe, Floerchinger:2009uf, Fukushima:2021ctq}. By now this technique is widely used for very different phenomena with emergent composites, ranging from condensed matter systems to Quantum Chromodynamics (QCD), for recent reviews and developments see  \cite{Dupuis:2020fhh, Fukushima:2021ctq, Fu:2022gou}. In contrast to effective field theories, the flows with emergent composites allow us to continuously interpolate between the physics described by different degrees of freedom. A prominent example is QCD, where flows with emergent composites take care of the dynamical formation of hadrons (dynamical hadronisation) at low momentum scales or temperatures. 

This work is organised as follows. In \Cref{sec:PC}, we introduce the details of PC and in particular, the model of FPC sought to generate the SM fermion masses. 
In \Cref{sec:EffAct}, we discuss the relevant terms of  the effective action of the fundamental theory, and in \Cref{sec:EffActDynHad} we present in detail the emergent composite approach for the composites of interest. In \Cref{sec:DimComposite}, we discuss the dimensionality of the composite and in \Cref{sec:FPCinEffAction}  derive the momentum scaling of the linear coupling of interest from the effective action. 
In \Cref{sec:Walkingregimes}, we discuss the role of the walking regime and estimate the properties of the bound states from the fundamental effective action. 
Subsequently, in \Cref{sec:fRG+MassHierachies} we provide a brief introduction to the fRG. The latter is used to compute the anomalous scaling of the relevant two-point functions, which allows us to derive an estimate for a particular minimal realisation of FPC. 
Finally, in \Cref{sec:Conc} we conclude. Many of the technical derivations and discussions are deferred to the Appendix.

%%%%%%%%%%%%%%%%%%%%%%%%%%%%%%%%%%%%%%%
\section{FPC and the effective action} 
\label{sec:PC}

Fundamental CH models are usually constructed as extensions of the SM by introducing a new confining gauge group $G_{\rm TC}$ and TC-charged chiral fermions ${\cal F}^{\alpha, a}$. As the TC-gauge coupling becomes strong towards the IR, it will give rise to the formation of composites. Given the TC-fermions transform non-trivially under a global `flavour' symmetry ${\cal G}$, the fermionic condensate 
\begin{equation}\label{eq:condensateCH}
\langle {\cal F}^{\alpha, a} \epsilon_{\rm TC}\, {\cal F}^{\alpha, b} \rangle
= \Lambda_c f^2 \Sigma_\theta^{ab}\,,
\end{equation}
spontaneously breaks the global symmetry ${\cal G} \to {\cal H}$, realizing the Higgs as a pseudo-Nambu-Goldstone boson (pNGB) of the coset ${\cal G}/{\cal H}$.
In \labelcref{eq:condensateCH}, $\alpha$ corresponds to the TC-gauge group index while $a$ and $b$ are flavour indices of the group ${\cal G}$. 
The latter contains the SM group $G_{\rm SM}$ to ensure a viable phenomenology. 
Moreover, $\Lambda_c \sim 4\pi f$ denotes the TC condensation scale with $f$ being the pNGB decay constant, $\epsilon_{\rm TC}$ is the antisymmetric tensor of $G_{\rm TC}$. Finally, $\Sigma_\theta$ is a matrix that parametrises the alignment of the unbroken symmetry in the physical vacuum, in particular with respect to the gauged subgroup of $\cal G$ (usually $G_{\rm SM}$). A misalignment then leads to EWSB.

After the global symmetry ${\cal G}$ is explicitly broken, the pNGBs obtain a vacuum expectation value which shifts the unbroken symmetry ${\cal H}$ by a finite amount such that it does no longer include $G_{\rm SM}$. 
The pNGBs $\Pi_{\hat a}$ of ${\cal G}/{\cal H}$ (including the Higgs boson) are parameterised as fluctuations around the vacuum via the Goldstone matrix
\begin{equation}
\Sigma(x)= \exp\left[ i \frac{2 \sqrt 2}{f}\Pi_{\hat a}(x) T_\theta^{\hat a}\right] \Sigma_\theta\,,
\end{equation}
where $T_\theta^{\hat a}$ are the broken generators of ${\cal G}/{\cal H}$. A common minimal choice in CH model building is $G_{\rm TC}= \text{Sp}(N_{\rm TC})$ with four Weyl fermions~${\cal F}^{\alpha, a}, a=1,..,4$, per techni-color~\cite{Cacciapaglia:2017cdi,Sannino:2017utc}, resulting in the global symmetry-breaking pattern SU$(4)_{\cal F}\to \text{Sp}(4)_{\cal F}$ after fermion condensation. This leads to ${\rm dim}[ \text{SU}(4)_{\cal F}/ \text{Sp}(4)_{\cal F}]\!=\!5$ pNGBs, with three of them, $\Pi_{1,2,3}$ being the electroweak Goldstone modes, $\Pi_4$ the Higgs boson, and $\Pi_5$ an additional electroweak singlet. In this case, the vacuum matrix reads
\begin{equation}
\Sigma_\theta = \cos\theta \begin{pmatrix}
i \sigma_2 & 0  \\
0 & -i \sigma_2
\end{pmatrix} + \sin\theta \begin{pmatrix}
0 & \bf{1}_2\\
-\bf{1}_2 & 0
\end{pmatrix}\,,
\end{equation}
with the electroweak vacuum expectation value $v=\sin \theta f$, such that $\sin\theta\!=\!0$  $(\sin\theta\!=\!1)$ corresponds to unbroken (fully broken) electroweak symmetry. We note that for the more minimal coset of ${\cal G}/{\cal H}= \text{SO}(5)/ \text{SO}(4)$, also preserving custodial symmetry, no fundamental four dimensional UV completion with elementary fermions exists~\cite{Cacciapaglia:2020kgq}.

The idea of PC provides a mechanism to reproduce the large hierarchies in the SM Yukawa couplings by introducing a linear mixing term between SM fermions and bound states of the composite sector. Focusing on the third-generation up-type quarks, the PC Lagrangian reads
\begin{equation}
\label{eq:PC}
{\cal L}_{\rm mix} = \frac{ \overline\lambda_q}{\Lambda_{\rm UV}^{\gamma_{\cal B}^q}}\, \overline q_L {\cal O}_{\cal B}^q + \frac{ \overline \lambda_t}{\Lambda_{\rm UV}^{\gamma_{\cal B}^t}}\, \overline t_R {\cal O}_{\cal B}^t  \, \, +{\rm h.c.}\,.
\end{equation} 
Here, $q_L$ and $t_R$ are the embeddings of the SM-like fields into irreducible representations of the global symmetry $\cal G$ of the composite sector. The couplings $ \overline \lambda_{q,t}$ are dimensionless ${\cal O}(1)$ parameters at the $ \Lambda_{\rm UV}$ scale and $\gamma_{\cal B}^{q}$ and $\gamma_{\cal B}^{t}$ are the respective anomalous dimensions of the composite-sector operators. As the TC-gauge coupling becomes strong, the composite operators can excite fermionic resonances
\begin{equation}
\langle 0| {\cal O}_{\cal B}^{q,t}| {\cal B}^{q,t} \rangle \neq 0\,,  
\end{equation}
which lead to the light fermion mass eigenstates becoming superpositions of the elementary fermions $q,t$ and the resonances. 

At an IR scale $\Lambda_c$, the linear mixing couplings will read as dictated by their renormalisation group (RG) scaling
\begin{align}
    \overline \lambda_q(\Lambda_{\textrm{c}})\simeq \overline \lambda_q(\Lambda_{\textrm{UV}})\, (\Lambda_c/\Lambda_{\rm UV})^{\gamma_{\cal B}^q}\,.
\end{align}
The large hierarchies between the SM Yukawa couplings are translated to the parameters $\bar \lambda_q$ and are naturally explained by their RG evolution over many orders of magnitude. 
While the linear mixing couplings are expected to be ${\cal O}(1)$ at $\Lambda_{\rm UV}$, at the IR scale $\Lambda_{\rm c}$ they will split up such as to correctly reproduce the SM fermion masses. 
In PC, both scales are largely separated to avoid an unachievable large scaling of the coupling. This is accomplished when the gauge coupling enters a quasi-fixed-point regime and freezes the TC-gluon dynamics. 
This is commonly referred to as a \textit{walking regime}. 

In summary, these two key ingredients together; an adequate scaling of the linear mixing coupling and the presence of the walking regime determine whether the SM Yukawa couplings can be reproduced. In the following, we will detail this mechanism for the case of FPC and derive in detail the scaling of the linear mixing operators from the effective action.

%%%%%%%%%%%%%%%%%%%%%%%%%%%%%%%%%%%%%%%%%%%%%%
\subsection{Effective action}
\label{sec:EffAct}

To realise FPC, fundamental scalars ${\cal S}^{\alpha, i}$ charged under the TC-gauge group are introduced in the fundamental CH action~\cite{Sannino:2016sfx,Cacciapaglia:2017cdi,Agugliaro:2019wtf,Cacciapaglia:2020kgq,FCDReview}. In the present renormalisation-group approach  based on the full effective action $\Gamma$, the strongly coupled TC part of $\Gamma$ is parameterised as 
\begin{align}\nonumber
\Gamma_{\rm CH}=&\,\int_x \Biggl\{ \frac{Z_A}{4} \,G_{\mu \nu} G_{\mu \nu} +\mathcal{L}_{\text{gf+ghosts}}  \\[1ex]\nonumber
& \hspace{-.9cm}+ \frac{Z_{{\cal S}}}{2}\left[ \left( D_\mu {\cal S}^{i}\right)^\dagger  \left( D_\mu {\cal S}^{i}\right) +  {{\cal S}^{i}}^{\dagger} m^2_{{\cal S}}\, {\cal S}^{i}\right] \\[1ex] \nonumber
&\hspace{-.9cm}+ Z_{{\cal F}}\,\bar{{\cal F}}^{a} \left(\sigma_\mu D_\mu + m_{{\cal F}} \right) {\cal F}^{a}\\[1ex] 
& \hspace{-.9cm}+ \sqrt{Z_{\psi}\,Z_{{\cal F}}\, Z_{{\cal S}}} \, y^{i,a}_{\rm TC} \,\psi^{i,a}\, \epsilon_{ij}\, \Phi^{j}\, \epsilon_{\rm TC}\, {\cal F}^a +{\rm h.c.} + \cdots\!\Biggr\},
\label{eq:CHEffectiveAction}
\end{align}
where the dots $\cdots$ in the last line stand for higher-order interaction terms in the fields. The TC-gauge field strength tensor reads 
\begin{align}
    G_{\mu \nu}= \partial_\mu A_\nu -\partial_\nu A_\mu  - \imag \, g_{\rm TC} \left[ A_\mu,  A_\nu\right]\,,
\end{align}
with the covariant derivative
\begin{align}
    D_\mu =\partial_\mu - \imag \, g_{\rm TC} A_\mu \,.
\end{align}
In \labelcref{eq:CHEffectiveAction} and from here on, we keep the colour indices and generators of the TC-gauge group implicit. 

In the following, we concentrate on the most minimal incarnation of $\text{SU}(4)_{\cal F}/\text{Sp}(4)_{\cal F}$ FPC. 
The TC-fermions are accordingly assumed to form a weak (chiral) doublet ${\cal F}^{1,2}$ with vanishing hypercharge as well as two SU$(2)_L$ singlets ${\cal F}^{3,4}$ with hypercharges $Y=\mp 1/2$ (per technicolour). 
Beyond that, the addition of 12 complex scalar degrees of freedom, residing in {\it 3 generations} of colour triplets ${\cal S}_q$ with hypercharge $Y=-1/6$, and corresponding colour singlets ${\cal S}_l$ with $Y=1/2$, allows for appropriate composite operators for all families of quarks and leptons. 
Both, TC-fermion and -scalar fields, transform under the fundamental representation of the TC-gauge group SU$(2)_{\rm TC}\sim$Sp$(2)_{\rm TC}$. This economic realisation of the FPC idea is accordingly called ``minimal fundamental partial compositeness" (MFPC)~\cite{Sannino:2016sfx,Cacciapaglia:2017cdi,Sannino:2017utc}. 
Further possible emerging coset structures in CH models have been widely investigated, both from an IR~\cite{Bellazzini:2014yua} and a UV~\cite{Belyaev:2016ftv,Ferretti:2016upr,Sannino:2016sfx} perspective.  
In the absence of fundamental fermion and scalar masses, $m_{\cal F}=m_{\cal S}=0$, the TC-fermions (scalars) exhibit a global SU$(4)_{\cal F}$ (Sp$(24)_{\cal S}$) flavour symmetry, corresponding to transformations along the index $a$ ($i$) in \labelcref{eq:CHEffectiveAction}. 
We note that, in the following, we will neglect the mass term for the TC-fermion fields and assume a flavour-trivial scalar mass term for simplicity.

The last line of \labelcref{eq:CHEffectiveAction} comprises the Yukawa interaction between the SM fermions (shown here as a spurion field $\psi^{i,a}$ of the full global symmetry) and the fundamental TC-fields. 
Here, $y^{i,a}_{\rm TC}$ are the Yukawa couplings associated to each component of the spurion field, $\epsilon_{ij}$ is the anti-symmetric tensor in Sp$(24)_{\cal S}$ and the TC-scalar fields have been arranged as $\Phi=({\cal S}, - \epsilon_{\rm TC} {\cal S}^\ast)^{\rm T}$, see \cite{Cacciapaglia:2020kgq,Sannino:2017utc,Sannino:2016sfx,FCDReview,Goertz2021}.  

We emphasise that the complete effective action $\Gamma$ and also its TC part~\labelcref{eq:CHEffectiveAction} should not be confused with their classical counterparts. 
The effective action $\Gamma$ takes into account the full quantum dynamics of the theory. We would like to elucidate this important aspect with the example of physical running couplings of given scattering processes. 
The respective scattering vertices or rather their one-particle irreducible (1PI) part are given by the nth derivative of the effective action w.r.t.~the fields involved in the scattering,  
\begin{align}
\Gamma^{(n)}_{\phi_{i_1}\cdots \phi_{i_n}}[\phi](p_1,...,p_n) =\frac{\delta}{\delta \phi_{i_1}(p_1)}\cdots \frac{\delta}{\delta\phi_{i_n}(p_n)}\Gamma[\phi]\,, 
\label{eq:DefGn}
\end{align}
where $\phi$ is a superfield that comprises all fields in the theory. 
In the present case it reads 
\begin{align}
\phi=( A_\mu, c,\bar c, {\cal S}, {\cal F}, \bar{\cal F},...)\,,
\label{eq:phif}
\end{align}
with the ghost and anti-ghost $c,\bar c$. 
In the following we shall also use the abbreviations $\Gamma^{(n)}$ and $\Gamma_{\phi_{i_1}\cdots \phi_{i_n}}$ for the sake of readability. 

The vertices \labelcref{eq:DefGn} are not renormalisation group invariant and carry the inverse RG-scaling of the fields. Their RG-invariant core is defined by 
\begin{align}
	\bar \Gamma^{(n)}_{\phi_{i_1}\cdots \phi_{i_n}}[\phi](p_1,...,p_n) =\frac{\Gamma^{(n)}_{\phi_{i_1}\cdots \phi_{i_n}}[\phi](p_1,...,p_n)}{\prod_{j=1}^n Z^{1/2}_{\phi_{{i_j}}}(p_j) } \,, 
	\label{eq:DefbarGn}
\end{align}
which is the non-perturbative version of the standard definition of running couplings in perturbation theory. The amputation of the external wave function reflects the amputation of external propagators in the LSZ-formalism and \labelcref{eq:DefbarGn} is simply the 1PI core of the respective scattering amplitude, for a recent discussion in QCD see \cite{Horak:2023xfb}. \Cref{eq:DefbarGn} carries the full momentum dependence of the given process, and hence directly provides the information about unitarity and momentum scaling. Finally, we note that $\bar\Gamma^{(n)}$ is simply obtained by derivatives w.r.t.~
\begin{align} 
Z_\phi^{1/2}(p)\, \phi(p)\,,
\label{eq:Z12phi}
\end{align}
which is the RG invariant field. 
\begin{figure}[t]
    \centering
    \vspace{-.3cm}
    \begin{align}
     \hspace{-0.5cm} \parbox{2. cm}{\centering\begin{tikzpicture}[scale=0.7]\input{./figures/4TCFermi.tex}\end{tikzpicture}}  \hspace{-0.5cm}&= \hspace{-0.2cm}\parbox{2.cm}{\centering\begin{tikzpicture}[scale=0.7]\input{./figures/4PFgluons.tex}\end{tikzpicture}}+\hspace{-0.2cm}\parbox{2.cm}{\centering\begin{tikzpicture}[scale=0.7]\input{./figures/4PFtriangle.tex}\end{tikzpicture}}+\hspace{-0.2cm}\parbox{2.cm}{\centering\begin{tikzpicture}[scale=0.7]\input{./figures/4PFSM.tex}\end{tikzpicture}} +\,\cdots \notag
    \end{align}
    \vspace{-.5cm}
    
    \caption{Generation of the interaction between two TC-fermions (plain black arrow line) and two TC-scalars (dashed) from the TC-gluons (curly) and SM fermions (plain blue arrow line) mediated box diagrams. 
    All vertices ($\Gamma^{(n)}(p_1,...,p_n)$, black and blue dots) and propagators ($1/\Gamma^{(2)}(p_1,p_2)$) are full (inverse) correlation functions. 
    The dots $\cdots$ represent additional loop diagrams simultaneously containing TC-gluon and SM fermion propagators and/or 4-field interactions. See text for details.}
    \label{Fig:FFSSgen}
\end{figure}
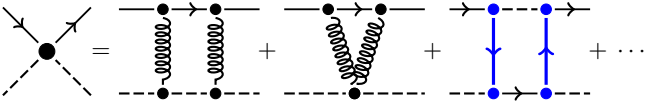

The effective action \labelcref{eq:CHEffectiveAction} also includes higher-order interaction terms of fundamental TC fields. 
These are encoded in the $\cdots$ and are  generated from loop diagrams. 
A relevant example is the two-scalar--two-fermion scattering, which is generated from the gauge-mediated box diagrams depicted in \Cref{Fig:FFSSgen}. 
This leads to terms in the full effective action $\Gamma$, or rather its technicolour part $\Gamma_{\rm CH}$, 
\begin{align}
    \Gamma_{{\cal S}^2{\cal F}^2 }=\int_x 
     Z_{{\cal S}} Z_{{\cal F}} \,g_{{\cal S}{\cal F}}\,{\cal S}^\dagger {\cal S}\,\bar{{\cal F}}{\cal F} +\cdots \,, 
\label{eq:4TCinteraction}
\end{align}
where $\cdots$ indicate all further two-scalar--two-fermion terms in the effective action.
Further, in \labelcref{eq:4TCinteraction} we have dropped the momentum dependence of the wave functions $Z_{{\cal S}}$ and $ Z_{{\cal F}}$ as well as that of the coupling $g_{{\cal S}{\cal F}}$.
From \labelcref{eq:4TCinteraction} we are lead to the momentum-dependent and RG-invariant scattering coupling as defined in \labelcref{eq:DefbarGn}, 
\begin{align} 
{\cal P}_{{\cal S}^\dagger {\cal S}\,\bar{{\cal F}}\,{\cal F}} \bar\Gamma^{(4)}_{{\cal S} {\cal S}^\dagger {\cal F}\bar{{\cal F}}}(p_1,p_2,p_3,p_4) \simeq 
g_{{\cal S}{\cal F}}(p_1,p_2,p_3) \,,
\label{eq:barGamma4}
\end{align}
where ${\cal P}_{{\cal S}^\dagger {\cal S}\,\bar{{\cal F}}{\cal F}}$ indicates the projection on the two-scalar--two-fermion term in \labelcref{eq:4TCinteraction}. 
In \labelcref{eq:barGamma4} we have omitted the momentum conservation $(2\pi)^4\delta(p_1+\cdots +p_4)$ as well as the Dirac, flavour and technicolour tensor structures as indicated by $\simeq$.  
While the discussion of a complete tensor basis for these terms goes beyond the scope of the present work, we provide a specific example with a vector tensor structure in the fermionic part: ${\cal S}^\dagger {\cal S}\,\bar{{\cal F}}\slashed{\partial}{\cal F}$. 
The respective RG-invariant scattering coupling $g^{(v)}_{{\cal S}{\cal F}}$ is given by 
\begin{align} 
{\cal P}_{{\cal S}^\dagger {\cal S}\,\bar{{\cal F}}\slashed{\partial}{\cal F}} \bar\Gamma^{(4)}_{{\cal S} {\cal S}^\dagger {\cal F}\bar{{\cal F}}}(p_1,p_2,p_3,p_4) \simeq 
g^{(v)}_{{\cal S}{\cal F}}(p_1,p_2,p_3)\, \slashed{p}_3 \,,
\label{eq:barGamma4pslashed}
\end{align}
where ${\cal P}_{{\cal S}^\dagger {\cal S}\,\bar{{\cal F}}\slashed{\partial}{\cal F}} $ indicates the projection on the term  ${\cal S}^\dagger {\cal S}\, \bar{{\cal F}}\slashed{\partial}{\cal F}$ in \labelcref{eq:4TCinteraction}. 

As can be inferred from \Cref{Fig:FFSSgen}, the scattering coupling strength is proportional to a combination of $g_\textrm{TC}^4$ and $y_\textrm{TC}^4$, multiplied with inverse powers of the momenta $p_i$, flowing through the diagram. 
For illustration, we restrict ourselves to a symmetric point with $p_i^2=p^2$. 
Then, the momentum scaling reads  
\begin{align}
    g_{{\cal S}{\cal F}}(p) \propto \frac{r_g \,g_\textrm{\tiny{TC}}^4(p) + r_y \,y_\textrm{TC}^4(p)}{\sqrt{p^2}}\,, 
\label{eq:gSF-g4}
\end{align}
where $r_{g/y}$ encode the combinatorial factors of each box diagram. Accordingly, \labelcref{eq:4TCinteraction} is suppressed in the ultraviolet due to the asymptotic freedom of the TC-gauge group and the Yukawa interaction. In turn, in the infrared, the TC-gauge coupling grows strong, and \labelcref{eq:4TCinteraction}, as well as further gauge-mediated interactions between the scalar and fermionic fields, become more relevant. 

We also note that the right-hand side of \Cref{Fig:FFSSgen} contains further diagrams, for example, a fish diagram with two two-scalar--two-fermion vertices and a mixed diagram with one TC-gluon exchange and a two-scalar--two-fermion vertex.  Moreover, further one-loop diagrams such as  box diagrams mixing TC-gauge fields and SM-fermions are also present. For small $g_{\textrm{\tiny{TC}}}\ll 1$ these diagrams are suppressed as they are proportional to $g_{\textrm{\tiny{TC}}}^8$ (fish diagram) and $g_{\textrm{\tiny{TC}}}^6$ (mixed diagram). 

%%%%%%%%%%%%%%%%%%%%%%%%%%%%%%%%%%%%%%%%%%%%%%
\subsection{Emergent composites}
\label{sec:EffActDynHad}

Analogously to the formation of mesons in QCD from resonant four-quark scatterings, resonant two-scalar--two-fermion scatterings may give rise to the formation of the fermionic composites ${\cal B} \sim {\cal S F}$. These resonant channels are then well-described by the propagation of a new degree of freedom, a composite operator consisting of both fundamental TC fields. The respective composite operator is given by 
\begin{align}
\mathcal{O}^{a,\,i }_B \sim \epsilon_{i\,j}\,{\cal S}^{\alpha,j}\,{\cal F}^{\alpha,a} {\cal T}  \,,
\label{eq:CompositeOperator}
\end{align}
and may also include higher-order terms. Here, $\mathcal{O}^{a,\,i }_B$ carries the chirality of the TC-fermion and transforms as a singlet under the confining gauge group. Moreover, all symmetries of the TC fields (containing the SM gauge symmetries) are encoded in the indices $a$ and $i$. The undetermined operator $\mathcal{T}$ projects on the spin indices of the TC fields. Its consideration makes explicit the generality in the symmetry structure of $\mathcal{O}_B$ as well as a potential choice of elaborated spin configurations for excited states.

The non-perturbative phenomenon of the formation of composites will be treated by functional renormalisation group (fRG) techniques, introduced in \Cref{sec:FRG}. With the adequate choice of external momentum, the effective four-field interaction can be exactly rewritten as the exchange of a composite of both fundamental external fields and a residual contribution. This is known in the fRG approach to QCD as \textit{dynamical hadronisation} and more generally it is taking into account \textit{emergent compositeness} \cite{Gies:2001nw, Pawlowski:2005xe, Floerchinger:2009uf, Fukushima:2021ctq}. Here we follow \cite{Pawlowski:2005xe, Fu:2019hdw, Fukushima:2021ctq}, where the composite fields are simply introduced on the level of the path integral via the respective current term, 
\begin{align} 
\exp\left\{\int_x J_{\cal B}\,{\cal B}(\varphi_f)\right\}\,, 
\label{eq:CompositeCurrent}
\end{align}
where ${\cal B}$ is related to \labelcref{eq:CompositeOperator}. In \labelcref{eq:CompositeCurrent}, the field $\varphi_f$ is the fundamental super field that is integrated over in the path integral. Accordingly, the introduction of the composite fields in terms of a current does not signal a reduction from the fundamental to an effective theory but rather the convenient reparametrisation of the fundamental theory in terms of emergent composite degrees of freedom. A well-known and well-studied example of such a reparametrisation is the two-particle irreducible effective action, which is obtained by introducing a current for the two-point function. 

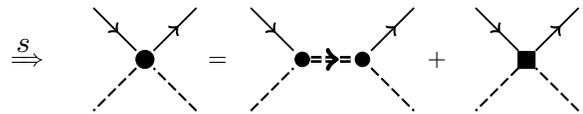
\begin{figure}[t]
    \centering
    \vspace{-.4cm}
    \begin{align}
 \xRightarrow{\scalebox{1.25}{$s$}}\hspace{0cm}\parbox{2.5 cm}{\centering\begin{tikzpicture}[scale=0.8]\input{./figures/4TCFermi.tex}\end{tikzpicture}} \hspace{-0.5cm}&= \parbox{2.5cm}{\centering\begin{tikzpicture}[scale=0.8]\input{./figures/4PFcomp.tex}\end{tikzpicture}} + \parbox{2 cm}{\centering\begin{tikzpicture}[scale=0.8]\input{./figures/4TCFermiRest.tex}\end{tikzpicture}}\notag
  \end{align}
    \vspace{-.7cm}
    \caption{Rewriting the full (fermion--scalar)$^2$ interaction in the $s$-channel as the exchange of the composite ${\cal B}$ plus a remnant interaction accounting for the momentum-dependence of the resonant tensor structure as well as the scattering in the other tensor structures. Double dashed lines with an arrow indicate the spin-1/2 composite propagators, plain lines with an arrow correspond to fundamental TC-fermion propagators and dashed lines to fundamental TC-scalar propagators.}
    \label{Fig:Bgeneration}
\end{figure}
Such a transformation allows us to rewrite a given scattering vertex in terms of an exchange of an emergent (resonant) composite field in a specific momentum channel and the remnant scattering, see \Cref{Fig:Bgeneration} for the composite $\cal B$ discussed above. At a given momentum scale and for bilinear composites this is reminiscent of a Hubbard-Stratonovich transformation. However, the fRG approach with flowing dynamical  composites~\cite{Gies:2001nw, Pawlowski:2005xe,Floerchinger:2009uf, Fu:2019hdw, Fukushima:2021ctq} also incorporates general composites~\cite{Pawlowski:2005xe, Fukushima:2021ctq}, not only bilinear ones. Further, it allows us to (re-)do this transformation at each scale in a mathematically well-defined way without potential overcounting problems. 

In the presence of the current term \labelcref{eq:CompositeCurrent} in the path integral, the respective effective action is obtained via a Legendre transformation and contains additional terms that carry the dynamics of the emergent composite field, 
\begin{align}\nonumber 
\Gamma_{{\cal B}}=& \int_x \Biggl\{ Z_{{\cal B}} \, \bar{{\cal B}}\, \left( \sigma_\mu \partial_\mu+ m_{{\cal B}}\right){\cal B} \\[1ex] 
& \hspace{-.7cm}+ h_{{\cal B}} \, \sqrt{Z_{{\cal B}}\,  Z_{{\cal S}}\,Z_{{\cal F}}} \left[{\cal S}\left(\bar{{\cal B}} {\cal F}  \right)+{\cal S}^\dagger \left(  \bar{{\cal F}}{\cal B} \right) \right]+\cdots\Biggr\}\,,
\label{eq:Baction}
\end{align}
where we have omitted flavour indices, chiral labels and the explicit consideration of $\mathcal{T}$ and the antisymmetric tensor $\epsilon_{ij}$. We emphasise that, as in \labelcref{eq:4TCinteraction}, all wave functions and the coupling $h_{\cal B}$ are momentum-dependent, and the latter is simply the RG-invariant scattering coupling 
\begin{align}\label{eq:Gamma3BSF}
\bar \Gamma^{(3)}_{{\cal S}{\cal F}\bar{{\cal B}}}(p_1,p_2,p_3) \simeq h_{\cal B}(p_1,p_2) \,,
\end{align}
where we omitted the momentum conservation $(2\pi)^4\delta(p_1+p_2+ p_3)$ as well as the Dirac, flavour and TC tensor structure. Importantly, \labelcref{eq:Baction} encodes the diagrammatic relation \Cref{Fig:Bgeneration} in terms of the equation of motion of the composite field ${\cal B}$. For $s$-channel momenta $s=(p_1+p_2)^2$ this relates \labelcref{eq:Baction} to \labelcref{eq:4TCinteraction} with 
\begin{align}
\left. \bar \Gamma_{{\cal B}}\right|_{ {\cal B}_{\textrm{\tiny{EoM}}}} = \bar \Gamma_{{\cal S}^2{\cal F}^2 }
\label{eq:GBGSF}
\end{align}
with the relation 
\begin{align} 
 \frac{h^2_{\cal B}(s)\,m_{\cal B}(s)}{s+m_{\cal B}^2(s)}\propto g_{{\cal S}{\cal F}}(s) \,, 
\end{align}
and $m_{\cal B}(s)\propto \sqrt{s}$. It follows with \labelcref{eq:gSF-g4} that $h^2_{\cal B}(s) m_{\cal B}(s)/(s+m_{\cal B}^2(s))\propto g_\textrm{\tiny{TC}}^4(s)/\sqrt{s}$, and hence decays rapidly in the UV. 
In turn, if the two-scalar--two-fermion $s$-channel scattering becomes resonant in the IR, this is well described in terms of the ${\cal B}$-exchange. Note, however, that the introduction of the emergent dynamical composite ${\cal B}$ only captures the respective momentum channel as well as further parts of the full scattering with $g_{{\cal S}{\cal F}}(p_1,p_2,p_3)$. 

We have mentioned below \labelcref{eq:CompositeOperator} that the field ${\cal B}$ or the operator ${\cal O}_{\cal B}$ may also include higher order terms. Indeed, \labelcref{eq:Baction} only comprises the lowest order terms in the composite field: a dispersion term for the composite state and two interaction terms which link fundamental fields' interactions to the propagation of a composite state. Higher order terms are generated as well and are indicated with $\cdots$. They contribute to the equations of motion of ${\cal B}$, leading to higher order terms in the fundamental fields and in particular ${\cal S}, {\cal F}$.

The full effective action in the presence of the composites is given by $\Gamma[\phi]$ and includes \labelcref{eq:CHEffectiveAction,eq:Baction}, where the superfield $\phi$ now also contains the composite field ${\cal B}$ and potentially further composites, 
\begin{align}
    \phi=(\phi_f,\phi_c)\,. 
\label{eq:phi}
\end{align}
The superfield $\phi_f$ is that of the fundamental fields, defined in \labelcref{eq:phif}. The superfield of all composites is denoted by $\phi_c$, to wit, 
\begin{align}
    \phi_f=( A_\mu, c,\bar c, {\cal S}, {\cal F}, \bar{\cal F},...)\,,\quad \phi_c=({\cal B}, \bar{\cal B},...)\,.
\label{eq:Superfield}
\end{align}
The dots in \labelcref{eq:Superfield} indicate that we may also introduce additional composite fields for further resonant interaction channels. 
This also entails the dynamics of such a system are well-captured by the present approach as long as all channels are considered: the emergence of a dynamical composite field is signalled by a divergent or resonant interaction channel. This resonant channel can be treated with the emergent compositeness approach discussed very generically in \cite{Fukushima:2021ctq}. 

We have already emphasised that the present approach describes the full theory. 
Its conceptual and technical advantage is the explicit appearance of the emergent dynamical degrees of freedom, that otherwise have to be described in terms of resonant interaction channels of higher-order correlation functions. 
The latter is technically more challenging and also has the disadvantage that higher-order scatterings of the emergent fields are far more difficult to describe in terms of the fundamental fields. 
An illuminating example is low energy QCD which is dominated by (multi) pion exchange processes. 
While these processes are complicated multi-quark scatterings, they are far simpler taken into account by multi-pion diagrams. This structure is at the root of chiral perturbation theory. 

Still, we can describe the theory in terms of the effective action of the fundamental degrees of freedom. 
The relation between the two effective actions $\Gamma[\phi_f]$ and $\Gamma[\phi]$ is given by 
\begin{align}
\Gamma[\phi_f]= \Gamma[\phi_f, \phi_{c}^{\textrm{\tiny{EoM}}}(\phi_f)]\,,
\label{eq:GcEoM=G}
\end{align}
where $\phi_{c}^{\textrm{\tiny{EoM}}}$ is the solution to the equations of motion (EoM) of the composites in the presence of a background $\phi_f$ of the fundamental fields, 
\begin{align}
\left. \frac{\delta \Gamma[\phi]}{\delta\phi_c}\right|_{\phi_c=\phi_{c}^{\textrm{\tiny{EoM}}}}=0\,.
\label{eq:EoMphic}
\end{align}
A solution of the EoM \labelcref{eq:EoMphic} for $\cal B$ is tantamount to $J_{\cal B}=0$, and hence \labelcref{eq:GcEoM=G}. 
Its lowest order version is given by \labelcref{eq:GBGSF}. 

\Cref{eq:GcEoM=G} highlights the fact that the introduction of the composites does not entail the reduction of the full theory to an effective one. 
To make this even more explicit, we close this discussion with the remark that the correlation functions of $\phi_f$ are given by 
\begin{align}\nonumber 
    \Gamma^{(n)}_{\phi_{f,i_1}\cdots \phi_{f,i_n}}[\phi_f]= &\,\frac{\delta\Gamma[\phi_f, \phi_{c}^{\textrm{\tiny{EoM}}}(\phi_f)]}{\delta \phi_{f,i_1}\cdots \phi_{f,i_n} }\\[1ex]& \hspace{-3cm}=\left.\prod_{j=1}^n\left[\frac{ \delta}{\delta \phi_{f,i_j}} + \frac{\delta \phi_{c,r}^{\textrm{\tiny{EoM}}}(\phi_f)}{\delta\phi_{f,i_j}} \frac{ \delta}{\delta \phi_{c,r}}\right] \Gamma[\phi_f,\phi_c]\right|_{\phi_c=\phi_{c}^{\textrm{\tiny{EoM}}}}\!\!\!.
    \label{eq:GfnGn}
\end{align}
\Cref{eq:GfnGn} entails, that a given correlation function of the fundamental fields is split into the part described by the exchange of the composites and the remaining part. For a detailed discussion of these parametrisations of correlation functions see \cite{Fu:2019hdw, Fukushima:2021ctq}. 

We close this Section with a short summary of the relevant properties and advantages of the present approach. It allows us to investigate the theory in a global and complete manner, taking into account fundamental and composite degrees of freedom simultaneously as well as their interplay. The effective action evolves from UV regimes dominated by the interplay of TC-gauge and TC-matter to a transition or interface regime where the rising gauge couplings lead to emergent dynamical composites. In the infrared below this transition regime, the dynamics are dominated by that of the emergent composites. The latter regime is typically described by means of low-energy effective field theories. Then, the respective low-energy couplings or parameters are tuned such that they accommodate the experimental low-energy measurements. In the present effective action approach, these couplings are determined consistently from those in the fundamental high energy theory, thus restricting the latter. The accuracy or rather predictive power of these constraints hinges on a quantitative understanding of the strongly correlated dynamics in the interface regime in which the composites emerge and become dynamical. These dynamics are well-captured by the present approach whose technical implementation within the fRG is discussed in \Cref{sec:FRG}.

%%%%%%%%%%%%%%%%%%%%%%%%%%%%%%%%%%%%%%%%%%%%%%
\subsection{Dimension of the composite}
\label{sec:DimComposite}

In this Section, we elaborate on the dimensionality of the composite and the convention used. To begin with, the effective action only depends on the dressed combination $Z^{1/2}_{\cal B}(p) \,{\cal B}(p)$. 
This is also reflected by the fact that the RG-invariant scattering vertices \labelcref{eq:DefbarGn}, are the RG-invariant building blocks of $S$-matrix elements as are obtained by derivatives w.r.t.~$Z^{1/2}_\phi(p)\, \phi$, see \labelcref{eq:Z12phi}. 
The RG-invariant field $Z^{1/2}_{\cal B}(p)\, {\cal B}(p)$ has the canonical dimension 3/2, which follows from the dispersion relation for fermionic fields (such as the hadron-like composite of interest), 
\begin{align}
\left[ Z^{1/2}_{\cal B}\, {\cal B}\right]_c=3/2 \,. 
\label{eq:dispersiondimension}
\end{align}
Evidently, observables are independent of the choice of the canonical dimension $\left[{\cal B}\right]_c$ of the field. 
Specifically, all choices for the canonical dimension of the composite field, and in particular the common choices $\left[{\cal B}\right]_c=3/2$ or $\left[{\cal B}\right]_c=5/2$, lead to the same RG-invariant (physical) vertices $\bar\Gamma^{(n)}$ defined in \labelcref{eq:DefbarGn}. 
Moreover, $S$-matrix elements are constructed directly from tree-level diagrams with the $\bar\Gamma^{(n)}$, and hence unitarity or its failure is derived from their momentum dependence. 

From the above discussion it also follows that a convenient choice is provided by a dimensionless wave-function renormalisation $Z_{\cal B}$: 
If we would consider the canonical dimension of the composite to be the sum of the fermion and scalar canonical dimensions, $\left[{\cal B}\right]_c = \left[{\cal F}\right]_c + \left[{\cal S}\right]_c = 5/2 $, the canonical dimension of $Z_{\cal B}(p)$ in \labelcref{eq:dispersiondimension} would be minus two, in summary leading to \labelcref{eq:dispersiondimension}. 
In fact, the former is the choice used for QCD with dynamical hadronisation as well as in low-energy effective theories of QCD. 
There, composite baryons are described by fermionic operators with the canonical dimension $3/2$, while composite mesons are described by bosonic operators with the canonical dimension 1. 
The respective dimension of the constituents is $9/2$ for three quarks $q^3$ or $3$ for a $q \bar q$ pair following from the dimension of the quarks $[q]_c=3/2$.

%%%%%%%%%%%%%%%%%%%%%%%%%%%%%%%%%%%%%%
\subsection{Linear mixing operator and anomalous scaling}
\label{sec:FPCinEffAction}

With the introduction of the dynamical composites in the previous Sections, the respective effective action also comprises all terms whose generation is not prohibited by symmetries. One example is the linear mixing term 
\begin{align}
\Gamma_{\rm mix}=&\int_x \Bigl\{\,\lambda^L_t  Z^{1/2}_q Z^{1/2}_{{\cal B}} \ \bar{q}_L {\cal B}^q_R\notag \\[1ex]
& \hspace{1.5cm}+\lambda^R_t Z^{1/2}_t Z^{1/2}_{{\cal B}} \ \bar{t}_R {\cal B}^t_L
+\text{ h.c.}\Bigr\}\,,
\label{eq:qBaction} 
\end{align}
key for the realisation of the SM fermion masses in FPC. In~\labelcref{eq:qBaction}, we have restricted ourselves to the linear mixing term with the third generation of quarks. 
Moreover, all wave functions in \labelcref{eq:qBaction} and the RG-invariant couplings or mixing masses $\lambda^{L/R}_t(p)$ are momentum-dependent. 

Terms like \labelcref{eq:qBaction} carry part of the quantum fluctuations of the scatterings of the fundamental fields. 
The latter correlations (or rather their 1PI part) are obtained by taking the $n$th field derivative of the effective action of the fundamental fields $\Gamma[\phi_f]$ in \labelcref{eq:GcEoM=G}, see \labelcref{eq:GfnGn}. 
Evidently, this correlation function can also be obtained by taking $\phi_f$-derivatives of $\Gamma[\phi_f,\phi_c^{\textrm{\tiny {EoM}}}(\phi_f)]$ on the right-hand side of \labelcref{eq:GcEoM=G}. 
In our case we only consider one composite structure, $\phi_c={\cal B}$. The solution of its EoM in the present simple approximation with the kinetic term in  \labelcref{eq:Baction} and the linear mixing term \labelcref{eq:qBaction} is given by  
\begin{align}
    {\cal B}^{\textrm{\tiny{EoM}}}=  \sqrt{\frac{Z_{\cal S} Z_{\cal F}}{Z_{\cal B}}} \frac{h^2_{\cal B}}{\sigma_\mu\partial_\mu+m_{\cal B}} {\cal S} {\cal F} +\sqrt{\frac{Z_{f}}{Z_{\cal B}}}\lambda_f f\,,
\label{eq:BEoM}
\end{align}
where $f$ is the respective SM fermion with adequate chirality. For constant ${\cal B}$ the composite is simply proportional to ${\cal S F}$ and a linear term $f$ in the SM fermion. When inserted on the composite effective action \labelcref{eq:qBaction} we  recover the fundamental Yukawa interaction in \labelcref{eq:CHEffectiveAction} and a mass term for the SM fermion field $f$. The latter term reflects already the mass-like character of the linear mixing couplings in \labelcref{eq:qBaction}.

We now elucidate the important relation between the couplings in the fundamental and composite actions within a relevant example: 
the fundamental Yukawa interaction in the last line of \labelcref{eq:CHEffectiveAction}. In the present approximation, its strength is given by the Yukawa couplings $y^{i,a}_{\rm TC}$ in the effective action $\Gamma[\phi_f]$. 
Hence it is obtained from the $\psi,{\cal S},{\cal F}$-derivative of $\Gamma[\phi_f]$. 
In turn, we can also obtain it from the $\psi,{\cal S},{\cal F}$-derivative of $\Gamma[\phi_f,{\cal B}^{\textrm{\tiny{EoM}}}(\phi_f)]$. 
However, the effective action in the presence of composites also contains a Yukawa term with a coupling $y^{i,a}_{\rm c, TC}\neq y^{i,a}_{\rm TC}$. 
The relation between $y^{f}_{\rm TC}$, $y^{f}_{\rm c, TC}$ and  $\lambda_f$ is obtained from the $\psi,{\cal S},{\cal F}$-derivative of \labelcref{eq:GcEoM=G}, using \labelcref{eq:BEoM} and \labelcref{eq:qBaction}, 
\begin{align}
  y^{f}_{\rm TC} = y^{f}_{\rm c, TC} -\lambda_f \,\frac{ h^2_{\cal B}\,  m_{\cal B}}{p^2+ m_{\cal B}^2}   \,.
\label{eq:ycy}
\end{align}
\Cref{eq:ycy} entails that the introduction of a composite redistributes contributions of the quantum fluctuations to a given correlation function. 
Moreover, only part of the quantum fluctuations is still carried by the quantum fluctuations of the fundamental fields and the rest is carried by the mixed correlations of composites and fundamental fields. 

We emphasise that while it seems to be suggestive to identify $y^{f}_{\rm TC}$ with $y^{f}_{\rm c, TC}$, it clearly amounts to a double-counting of fluctuations as is evident from \labelcref{eq:ycy}. 
Importantly, in the present emergent composite approach, this double counting problem is absent as the correct relations between the different couplings such as \labelcref{eq:ycy} are implemented by definition. We further elaborate on \labelcref{eq:ycy} and the distribution of fluctuations in \Cref{sec:AnomalousDimCompositeSector}.

The canonical momentum dimension of any arbitrary fermion $f$ linear mixing coupling $\lambda_f$ is one, as follows from \labelcref{eq:ycy}. It is also easily inferred from the canonical dimension of the dressed fields, 
\begin{align}
\left[\sqrt{Z_{f}Z_{\cal B}}\, \bar{f}\, {\cal B}\right]_c=3\,.
\end{align}
which follows readily from the dimensional analysis of the kinetic terms of quarks $f$ and composite $\cal B$. Hence, the canonical dimension of $\lambda_f$ is given by $[\lambda_f]_c=1$. 

We have already discussed that all parameters in the effective action are momentum-dependent, which comprises the information about unitarity as well as the evolution of masses and couplings in the theory. 
If all parameters evolve according to their canonical momentum dimension, the theory is in the scaling regime of a fixed point. 
For the mass mixing parameter, this entails $\lambda_f(k) \propto k$, where $k$ is an average momentum scale. 

The anomalous momentum scaling of couplings and masses is best accessed from the scaling of their dimensionless counterparts. 
In the case of $\lambda_f$, the dimensionless coupling $\bar \lambda_f$ reads  
\begin{align}\label{eq:dimlesslambda}
 \overline \lambda_f&= \lambda_f \, k^{-1}\,.
\end{align}
The average momentum scale $k$ is also used as a cutoff scale in the fRG approach, employed for explicit computations in the present work. This is detailed in \Cref{sec:FRG}. 

Here we note that in the absence of an anomalous scaling, the dimensionless couplings and parameters do not scale with $k$. Conversely, the anomalous dimension is given by 
\begin{align}
\partial_t \overline \lambda_f&= \gamma_{\lambda_f} \, \overline \lambda_f\,,
\label{eq:gammalambdaq}
\end{align}
where $\partial_t \equiv k \, \partial_k$. We also note that the anomalous scaling of the wave functions $Z_\phi$ is given by 
\begin{align}
\partial_t Z_\phi =  \gamma_\phi \,Z_\phi\,.
    \label{eq:gammaphi}
\end{align}
\Cref{eq:gammaphi} entails that the anomalous scaling of the fields is given by $-\gamma_\phi$. 
For that reason, the common definition in fRG applications contains a minus sign in~\labelcref{eq:gammaphi}. 

We define the full momentum dimension $[{\cal O}]$ of an operator $\cal O$ with its $k$ scaling. 
For the mixed mass term~\labelcref{eq:qBaction} it follows that  
\begin{align}\label{eq:gammaqB}
\left[\overline \lambda_f \,k \,Z_{f}^{1/2}  Z_{\cal B}^{1/2}\right] =\gamma_{f{\cal B}}\,.
\end{align}
Then, the full dimension of the mixing coefficient $\overline\lambda_f$ is obtained by subtracting the scaling of the external fields and the canonical dimension carried in the explicit momentum dependence in $k$,
\begin{align}\label{eq:gammalambda}
\left[\overline\lambda_f\right]&\equiv \gamma^{\ }_{\lambda_f}= \gamma_{f\cal B} -1  -\frac{\gamma_f}{2} -\frac{\gamma_{\cal B}}{2}\,.
\end{align}
Moreover, given $\bar\lambda_f$ at a UV scale $k=\Lambda_\textrm{UV}$, its value at a measurable IR scale $k=\Lambda_\textrm{c}$ is given by the momentum scaling
\begin{align}\label{eq:scalinglambdaLambdaUVLambdac}
\overline \lambda_f (\Lambda_{c})&= \overline \lambda_f(\Lambda_\textrm{UV}) \left(\frac{\Lambda_\textrm{c}}{\Lambda_\textrm{UV}}\right)^{\gamma_{\lambda_f} }\,.
\end{align}
The function $\gamma_{\lambda_f}$ encloses all non-perturbative information on the composite and its dynamics. \Cref{eq:gammalambda} is agnostic to the particular composition of the fermionic bound-state ${\cal B}$ and hence holds for any type of fermionic composites (eg. ${\cal B} \sim {\cal F}{\cal F}{\cal F}$). Different natures of ${\cal B}$ will give rise to different quantum corrections in each of the scaling functions in \labelcref{eq:gammalambda}. 

Considering $\gamma_{f}$ to be negligible, the magnitude and sign of $\gamma_{\cal B}$ and $\gamma_{f \cal B}$ will lead to a more or less enhanced scaling of the coupling $\overline \lambda_f$. In the case where $\gamma_{f \cal B} - \gamma_{ \cal B}/2 \lesssim 0$ the coupling $\overline \lambda_f$ scales as that of a relevant operator (eg. as a fermionic mass term) and grows towards the IR. If $\gamma_{f\cal B}-\gamma_{\cal B}/2 \approx 1$, the scaling is like for a marginal operator and will show a conformal scaling at the fixed point if $\gamma_{\lambda_f}=0$. Last, if $\gamma_{\lambda_f}>0$, the coupling scales as an irrelevant operator growing towards the UV and decreasing towards the IR. This is explicitly necessary in the FPC framework in order to generate the masses of the lightest SM fermion generations.

We close this section by introducing some useful notations for the upcoming sections. It is instructive to define the \textit{degree of compositeness} at a scale $\Lambda_\textrm{c}$ (following the notation of~\cite{Goertz2021,Goertz:2021xlx}), which parametrises the strength of the mixing of the SM-like fermions with the composite resonances below the condensation scale 
\begin{align}\label{eq:PCeps}
\epsilon^{L}_f \equiv \frac{\overline{\lambda}^{L}_f\left(\Lambda_{\textrm{c}}\right)}{g_{\cal B}\left(\Lambda_{\textrm{c}}\right)} = \frac{\overline{\lambda}^{L}_f\left(\Lambda_{\textrm{UV}}\right)}{g_{\cal B}\left(\Lambda_{\textrm{c}}\right)}\left( \frac{\Lambda_{\textrm{c}}}{\Lambda_{\textrm{UV}}}\right)^{\gamma^{ }_{\lambda_f^L}}\,.
\end{align}
Here, $g_{\cal B}$ is the coupling strength of the resonances and $\Lambda_{\textrm{c}}$ is the IR condensation scale, reasonably expected around $\sim 10$\,TeV. For naturalness reasons, the dimensionless coupling $\overline \lambda^{L}_f (\Lambda_\textrm{UV})$ is expected to be of ${\cal O}(1)$.

Generalising to three generations of quarks and leptons, the hierarchical pattern of SM fermion masses (and mixings) can now be argued to arise from small differences in the anomalous dimensions $\gamma_{\lambda}$ via RG flow between the largely separated scales $\Lambda_{\textrm{c}}$ and $\Lambda_{\textrm{UV}}$. In fact, the resulting hierarchical degrees of compositeness enter the fermion masses, after integrating out the heavy resonances (see, e.g.~\cite{Goertz2021,Goertz:2021xlx,Panico:2015jxa}), which read
\begin{align}\label{eq:m}
m_f \sim \ g_{\cal B}\left(\Lambda_{\textrm{c}}\right) \frac{v}{\sqrt{2}} \epsilon^{L}_f \epsilon^{R}_f = \frac{\mathcal{Y}_f \,v}{\sqrt{2}}\left( \frac{\Lambda_{\textrm{c}}}{\Lambda_{\textrm{UV}}}\right)^{\!\left(\gamma^{\ }_{\lambda^{L}_f}+\gamma^{\ }_{\lambda^{R}_f}\right)}
\,,
\end{align} 
with
\begin{align}
\label{eq:Ynew}
\mathcal{Y}_f=& \overline{\lambda}^{L}_f (\Lambda_{\textrm{UV}}) \, \overline{\lambda}^{R}_f (\Lambda_{\textrm{UV}}) /g_{\cal B}\left(\Lambda_{\textrm{c}}\right) \,,
\end{align} 
which gathers all ${\cal O}(1)$ quantities.

%%%%%%%%%%%%%%%%%%%%%%%%%%%%%%%%%%%%%%%%%%%
\subsection{Emergent composites in the walking regime}\label{sec:Walkingregimes}

In FPC, a walking regime is expected to emerge around a scale $\Lambda_{\textrm{UV}}$ at which the beta function of the TC gauge coupling vanishes. As discussed in the previous Sections, this is required to generate the SM fermion masses through the momentum scaling of $\lambda_q$ along many orders of magnitude, see \labelcref{eq:scalinglambdaLambdaUVLambdac}. 

Walking regimes are generated from IR quasi-fixed-point solutions in the TC-gauge coupling. These are purely determined by the structure of the gauge group and the charged matter content. Depending on the number of fields and the representation under which they transform, this conformal scaling regime arises at different gauge coupling values and is related to Banks-Zaks (BZ) fixed points. The properties of the conformal window, such as the magnitude of the fixed point coupling and its boundaries, have been studied with perturbative methods \cite{Banks:1981nn,Dietrich:2006cm,Pica:2010xq}, and non-perturbative functional methods \cite{Braun:2006jd,Gies:2005as} as well as lattice simulations \cite{Ryttov:2009yw, Ryttov:2007sr}.

In FPC, dynamical chiral symmetry breaking is required to take place in order to form pNGBs which realise the Higgs boson. 
Through this mechanism, the TC-fermions become massive and decouple from the contributions to the TC-gauge beta function leading to an IR Landau pole and terminating the walking regime~\cite{Dietrich:2009ix,Diehl:2009ma,Dietrich:2010yw}. 
Therefore, it is argued \cite{Sannino:2016sfx,Cacciapaglia:2020kgq} that the different fundamental incarnations should be found outside (below) the conformal BZ window \cite{Pica:2010xq}.

In this work, we do not investigate further the existence of viable walking regimes within different FPC models but assume the scenario in which dynamical chiral symmetry breaking triggers at the end of the walking regime. While we delegate the analysis of the viable models with the fRG to an independent work, here we focus on the qualitative properties of the bound states of interest along the walking regime. 

In order to address whether the scaling $\gamma_{\lambda}$ reproduces the correct SM fermion masses, the parameters of the composite sector along the quasi-conformal regime need to be analysed. By rewriting the four-TC-field interaction term in \labelcref{eq:GBGSF} as the exchange of a composite, the Yukawa coupling $h_{\cal B}$ and the mass of the composite $m_{\cal B}$ in \labelcref{eq:Baction} can be estimated from the quantum corrections giving rise to the four TC-field interaction. In the case at hand, the leading quantum corrections are provided by the diagrams shown in \Cref{Fig:FFSSgen}.

The first box diagram is mediated by TC-gluons and will dominate as the gauge coupling enters a strong walking regime, allowing the formation of bound states and later on, triggering chiral symmetry breaking. 
Following, the Yukawa coupling between TC fields and SM fermions may become strong due to the effect of TC-gauge quantum corrections. 
In this case, the second SM fermion-mediated box diagram may contribute substantially to the formation of the bound state. 
The TC-gauge corrections on $y_{\rm TC}$ are subject to the group structure and the matter content in the theory and may lead to different qualitative results depending on the model~\cite{Sannino:2016sfx,Cacciapaglia:2020kgq}. 
One  particularly interesting scenario is where the formation of bound states is dominated by the exchange of SM fermions. In this case, $y_{\rm TC}\gg g_{\rm TC}$ via the TC-gluon enhancement in the quantum corrections to the Yukawa coupling between SM and TC fields. 
This scenario would help to relax the magnitude of the fixed point value $g^*_{\rm TC}$, reaching perturbative values, while still providing the formation of composites, purely dominated by Yukawa interactions.

In the walking regime, the momentum scaling of the different couplings is given by their canonical momentum dimension. Marginal couplings and dimensionless parameters (as the composite mass $\bar m_{\cal B}$) are hence fixed to a constant value, leading to the relation
\begin{align}
 \frac{(h^*_{\cal B})^2\,\bar m^*_{\cal B}}{1+  (\bar m_{\cal B}^* )^2}  \propto  r_g\left(g^*_\textrm{TC}\right)^4 +r_y \left(y^*_\textrm{TC}\right)^4\,.
\label{eq:hBapproxCW}
\end{align}
Here, $r_{g/y}$ encodes the diagrammatic combinatorial factors as well as the TC-color and spin structures of the respective box diagrams. This simplified relation allows us to estimate the composite sector parameters from the couplings of the fundamental theory.

It is important to stress that in principle $\bar m_{\cal B}$ and $h_{\cal B}$ could be precisely determined from the flow of 2- and 3-point functions of composites and fundamentals. However, this is beyond the scope of the current analysis and here we just estimate them via the relation \eqref{eq:hBapproxCW}, whose validity is backed up by knowledge from 2+1 flavours QCD~\cite{Mitter:2014wpa, Braun:2014ata, Cyrol:2017ewj, Fu:2019hdw}. The equivalent $h_{\cal B}$ for pions is known to be approximately constant even though the gauge coupling runs strongly. The respective coupling to QCD-composites shows an IR attractive quasi-fixed point behaviour in which the only variation is sourced by chiral symmetry breaking. This is caused by all the momentum scaling of $g_{\rm TC}$ being translated to the momentum scaling of the composite mass. In fact, this is the dynamical explanation for the emergence of chiral symmetry breaking. The strong interactions drive the meson masses to zero triggering spontaneous symmetry breaking of the chiral potential, see~\cite{Mitter:2014wpa,Braun:2014ata,Cyrol:2017ewj,Fu:2019hdw} for more details. In the current scenario, the gauge coupling reaches a fixed point and hence no variation is propagated to the composite parameters further supporting the estimate in \labelcref{eq:hBapproxCW}. 

%%%%%%%%%%%%%%%%%%%%%%%%%%%%%%%%%%%%%%%%
\section{Anomalous scaling and flavour hierarchies}\label{sec:fRG+MassHierachies}

In the present Section, we compute the anomalous scaling of different 2-point functions necessary to determine whether a successful reproduction of the  fermion mass hierarchies of the Standard Model particles is achievable within MFPC.

The fluctuation analysis of dynamics of the composite (as well as that of the other fields) is done within the functional renormalisation group approach, and in \Cref{sec:FRG} we briefly summarise the fRG approach to theories with emergent composites. This approach is used in \Cref{sec:AnomalousDimCompositeSector} to compute the anomalous dimensions of the composite sector, and as an example, we provide an estimate for a particular scenario. This is done on a conceptual level and a more quantitative study will be presented elsewhere. 

%%%%%%%%%%%%%%%%%%%%%%%%%%%%%%%%%%%%%%%%%%%%%%%%%
\subsection{Functional Renormalisation Group }\label{sec:FRG}

For the computation of the momentum scaling of the linear mixing coupling $\lambda_f$ and future analysis of the dynamics of CH sectors, we will employ the functional renormalisation group \cite{Wetterich:1991be,Morris:1993qb,Ellwanger:1993mw}, which already has proven its applicability in very similar strongly coupled regimes. 
For a general review see \cite{Dupuis:2020fhh}, for IR QCD studies see e.g.~\cite{Gies:2002hq, Braun:2014ata, Cyrol:2017ewj, Fu:2019hdw} and for SM applications \cite{Pastor-Gutierrez:2022nki, Gies:2013fua,Gies:2014xha,Eichhorn:2015kea,Held:2018cxd}. In this approach, an IR regulator quadratic in the fields is introduced at the level of the classical action,
\begin{align}
\Delta S _k [\phi]=\frac12  \int_x \, \phi (-q) \,R_k(q) \,\phi (q)\,. 
\label{eq:DeltaSk}
\end{align} 
\Cref{eq:DeltaSk} implements the Wilsonian idea of progressive integration of momentum shells by suppressing quantum fluctuations of momenta $p$ below a given IR cutoff scale~$k$. $R_k(q)$ is a matrix in field space and introduces mass-like terms for the different fields $\phi_i$ in the infrared. In turn, for large momenta $R_k(q)$ decays rapidly.   

The regulator insertion leads to the definition of an RG-cutoff dependent effective action $\Gamma_{k}[\phi]$. The full effective action is obtained after all quantum fluctuations have been integrated out at $k=0$ with $\Gamma[\phi]=\Gamma_{k\to 0}[\phi]$. The cutoff scale dependence of $\Gamma_{k}[\phi]$ is described by the Wetterich equation \cite{Wetterich:1992yh}, 
\begin{align}
	\partial_t \Gamma_k [\phi ] &= \frac{1}{2} \text{Tr}\,\left[\left( \frac{1}{\Gamma^{(2)}_k +R_k }\right)_{ij}\, \partial_t R^{ij}_k \right]\, ,
\label{eq:Flow}
\end{align}
where $\Gamma^{(2)}_k$ is the second functional derivative of the scale-dependent effective action $\Gamma_k$ w.r.t.~the fields $\phi$, see \labelcref{eq:DefGn}. The subscript ${}_{ij}$ indicates the different species $\phi_i$. Note that the regulator matrix $R^{ij}_k$ is diagonal in its bosonic components and symplectic in its fermionic components.  The derivative $\partial_t = k \,\partial_k$ is the logarithmic scale derivative w.r.t.~the infrared cutoff scale, which may be interpreted as an average momentum scale. 

The non-perturbative flow equation \labelcref{eq:Flow} is a one-loop exact equation that gives us access to the non-trivial strongly-coupled sector in CH scenarios, while remaining versatile to investigate large theory and parameter space.
Moreover, the dynamical emergence of bound states or more generally dynamical composites can be incorporated within a modified version of \labelcref{eq:Flow}, \cite{Gies:2001nw, Pawlowski:2005xe,Floerchinger:2009uf, Fu:2019hdw, Fukushima:2021ctq}. This allows us to consider the effective action in the presence of emergent composites in a mathematically sound way. This approach is known as \textit{dynamical condensation} or the fRG approach with \textit{emergent composites} \cite{Fukushima:2021ctq}, the latter literally describing its core. It has been widely employed in the areas of condensed matter and QCD. In the latter, it is commonly called \textit{dynamical hadronisation}.  

Its general version has been derived in \cite{Pawlowski:2005xe}, see also \cite{Wegner_1974} for a formulation for the Wilsonian effective action. In \cite{Fukushima:2021ctq} its application to a successive tower of composites is explained in the example of QCD (scalar-pseudoscalar mesons, vector mesons, diquarks and baryons), and this application is similar to the present use. Here we briefly discuss the generalisation in \cite{Pawlowski:2005xe} at the example of one emergent scale-dependent composite at lower momentum scales. While $\phi_c$ is an independent emergent field in the effective action, its flow is described by a functional  
\begin{align}
    \dot\phi_c[\phi]\,,
    \label{eq:dotphi}
\end{align}
with $\phi=(\phi_f,\phi_c)$ in \labelcref{eq:phi}. \Cref{eq:dotphi} takes care of the rotation of the field basis due to the scale dependence of the composite. Typically it is chosen such that the entire dynamics of the respective resonant channel is stored in the propagation of the composite. This choice optimises the field basis and hence stabilises expansion schemes of the effective action. In short, it leads to the most rapid convergence of the diagrammatic expansion of the full theory. 

In the presence of the composite, the flow \labelcref{eq:Flow} is modified by additional terms that take care of the scale-dependence of the emergent dynamical degree of freedom $\phi_c$. It reads, 
\begin{align}\nonumber 
	\left(\partial_t + \int \dot\phi_c \frac{\delta}{\delta\phi_c}\right)\,\Gamma_k [\phi]& \\[1ex] 
 &\hspace{-3.6cm}= \frac{1}{2} \text{Tr}\,\left[\left( \frac{1}{\Gamma_k^{(2)} +R_k }\right)_{ij}\, \left(\partial_t \delta^{jn}   +2 \frac{\delta \dot\phi_n }{\delta \phi_j} \right) R^{ni}_k \right]\, ,
\label{eq:GeneralFlow}
\end{align}
where $\dot\phi_i= \delta_{ic}\dot\phi_c$ in a slight abuse of notation. The additional term on the left-hand side of \labelcref{eq:GeneralFlow} simply accounts for the (field-dependent) anomalous dimension of the emergent composite, while the additional term on the right-hand side accounts for the rotation of the composite field axis in the field basis of the fundamental fields.

%%%%%%%%%%%%%%%%%%%%%%%%%%%%%%%%
\subsection{Anomalous dimensions}
\label{sec:AnomalousDimCompositeSector}

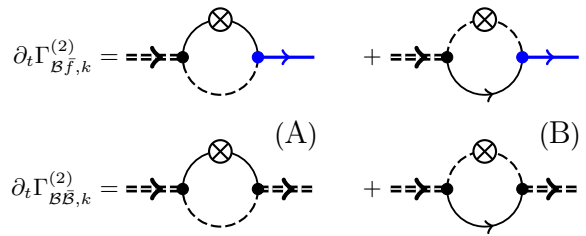
\begin{figure}
\centering
\begin{align}
\Large{\partial_t\Gamma^{\left(2\right) }_{{\cal B}\bar{f}, k} = } &\, \parbox{2.cm}{\centering\begin{tikzpicture}[scale=0.5]\input{./figures/BPsiBar_fermionreg.tex}\end{tikzpicture}}&&+ \parbox{2.cm}{\centering\begin{tikzpicture}[scale=0.5]\input{./figures/BPsiBar_scalarreg.tex}\end{tikzpicture}} \,\notag\\
\Large{\partial_t \Gamma^{\left(2\right) }_{{\cal B}\bar{\cal B}, k} = } &\, \parbox{2.cm}{\centering\begin{tikzpicture}[scale=0.5]\input{./figures/BBar_fermionreg.tex}\end{tikzpicture}}^{\big(\textrm{\large A}\big)} &&+ \parbox{2.cm}{\centering\begin{tikzpicture}[scale=0.5]\input{./figures/BBar_scalarreg.tex}\end{tikzpicture}}^{\big(\textrm{\large B}\big)}\,\notag
\end{align}
\vspace{-0.6cm}
\caption{Diagrammatic flow of the composite--SM-fermion two-point function (first line) and the composite's two-point function (second line), necessary for the computation of the vertex flow \labelcref{eq:gammafB} and the composite's anomalous dimension \labelcref{eq:etaB}. All vertices (black circles) and regulated propagators are full correlation functions. Double dashed lines with an arrow correspond to the composite regulated propagator, plain lines with an arrow correspond to fundamental TC-fermion regulated propagators, dashed lines to fundamental TC-scalar regulated propagators and blue lines with an arrow to the SM fermion regulated propagators. Crossed circles depict the insertion of $\partial_t R^{(\phi)}_k$. 
}
\label{fig:twopointB}
\end{figure}

We have discussed in \Cref{sec:PC} that the presence of a walking regime and an adequate scaling of the coupling between composites and SM fermions is necessary for reproducing the SM masses via FPC. In this Section, we compute the momentum scaling of an arbitrary linear mixing coupling $\bar\lambda_f$ between a SM-fermion and a composite field ${\cal B}$. Following from the derivation in \labelcref{eq:gammalambda}, the momentum scaling of the coupling reads
\begin{align}\label{eq:gammalambdasimplified}
\gamma_\lambda=-1+\gamma_{f{\cal B}}-\gamma_{f}/2-\gamma_{{\cal B}}/2 \,.
\end{align}
As discussed in \Cref{sec:Walkingregimes}, the magnitude of the fundamental Yukawa couplings $y^{*}_{\rm TC}$ is considered to be subdominant with respect to the gauge dynamics. This restricts ourselves to the QCD-like scenario in which the composite dynamics is driven by the TC-gauge dynamics. In this scenario, no sizeable contributions aside from the SM ones are present in $\gamma_f$. As these are known to be small and negligible \cite{Pastor-Gutierrez:2022nki}, the SM fermions anomalous dimension can be approximated as 
\begin{align}\label{eq:gammafto0}
\gamma_{f} \approx 0\,.
\end{align}
Nonetheless, it is important to note that scenarios where the Yukawa coupling becomes strong along with the TC-gauge coupling may lead to further improtant corrections to the SM fermion phenomenology,coupling, may lead to significant corrections on the SM which would be interesting to investigate further.

The momentum scaling of the vertex \labelcref{eq:gammaqB} and the anomalous dimension of the composite $\gamma_{{\cal B}}$ remain to be computed. We obtain the first from the linear-mixing 2-point function
\begin{align}\label{eq:two-pointBq}
\Gamma^{\left(2\right) }_{{\cal B}\bar{f}, k}= \frac{\delta^2 \Gamma_k}{\delta {\cal B}(p)\delta {\bar{ f}}(-p)}&= \!\lambda_f  \sqrt{Z_f Z_{{\cal B}}}\,,
\end{align}
by performing two functional derivatives of the full effective action (see \labelcref{eq:DefGn}) with respect to $\bar{f}$ and ${\cal B}$. By applying a scale derivative on both sides of \labelcref{eq:two-pointBq}, we obtain
\begin{align} \label{eq:gammafB}
\gamma_{f{\cal B}}&=   \,\partial_t\Gamma^{\left(2\right) }_{{\cal B}\bar{f}, k}\, /\,\Gamma^{\left(2\right) }_{{\cal B}\bar{f}, k}\,.
\end{align}
In the first line of \Cref{fig:twopointB}, the diagrammatic flow $\partial_t\Gamma^{\left(2\right) }_{{\cal B}\bar{f}, k}$ is depicted. As the linear mixing couplings carry a scalar-tensor structure (like a fermionic mass term) and there is no vector structure in any of the Yukawa couplings, the only non-vanishing contribution in the loop diagrams will be proportional to the scalar mode of the TC-fermion propagators. However, in \Cref{sec:FPCinEffAction} we argued vanishing $m_{\cal F}$ and $m_{\cal S}$ in order to preserve the global flavour symmetries. Additionally, TC masses will trigger the decoupling of the TC fields from the quantum corrections to the TC gauge coupling and hence invalidate the possibility of a walking regime. For this reason, the fundamental masses are taken to be $m_{\cal F},m_{\cal S}\leq\Lambda_{\rm c}$. With $m_{\cal F}=0$, each diagram vanishes, and we are led to 
\begin{align}\label{eq:gammaqB0}
\gamma_{f{\cal B}}&= 0\,.
\end{align}
It follows from this result that the linear mixing operators in \labelcref{eq:gammafB} are not generated by quantum fluctuations from the fundamental theory in the absence of a fundamental TC-fermion mass. 
However, we have discussed around \labelcref{eq:ycy} that the fundamental Yukawa coupling $y^f_{\rm TC}$ in the absence of the composites are only related but not equal to $ y^f_{\rm c, TC}$ in the presence of composites. 
The former is known to show non-vanishing quantum contributions $\partial_t y^f_{\rm TC}\neq 0$ in the fundamental theory, see e.g.~\cite{Sannino:2016sfx,Cacciapaglia:2020kgq}. 
These contributions are not reproduced by $\gamma_{f{\cal B}}$ in the current approximation but accounted in $\partial_t y^f_{\rm c, TC}$. On the other hand, contributions of $\partial_t y^f_{\rm TC}$ in the chirally broken phase are reproduced in $\gamma_{f{\cal B}}$ as reflected by its proportionality to $m_{\cal F}$.

\begin{figure*}[th]
    \includegraphics[width=2.\columnwidth]{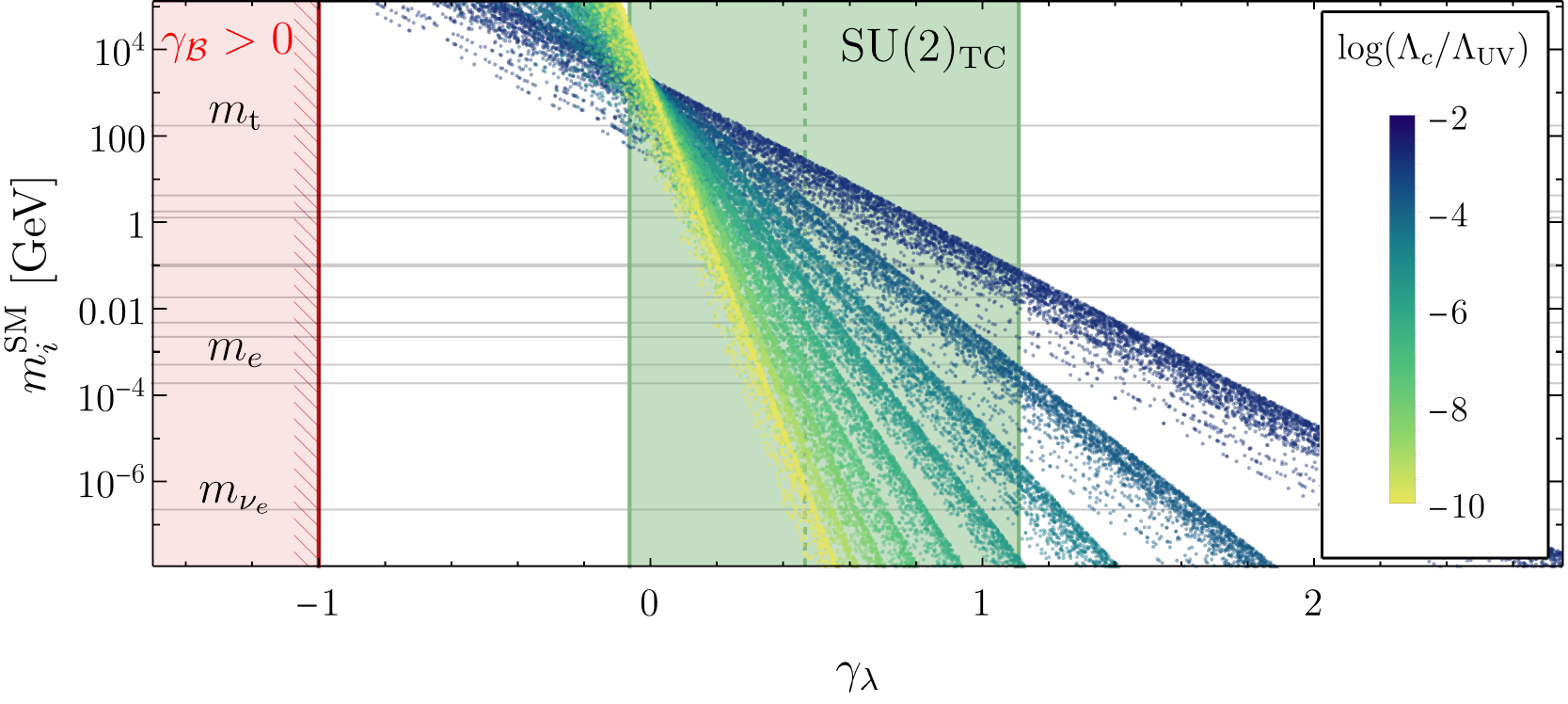}
    \caption{SM fermion masses as a function of the anomalous scaling $\gamma_\lambda$ defined in \labelcref{eq:gammalambdasimplified}. The colourful clouds of points show a parameter investigation in which ${\cal Y}_f\in  [0.1,4\pi]$ and $\gamma_\lambda \in [-1,3]$. From blue to yellow, the size of the walking regimes is discretely increased by varying $\Delta\Lambda=\log\left(\Lambda_{\textrm{c}}/\Lambda_{\textrm{UV}}\right)$ from $\Delta\Lambda=-2$ to $\Delta\Lambda=-10$. The red-shaded region depicts the prohibited area inaccessible due to the found negativity of the composite's anomalous dimension and coincidentally to the unitarity bound of the coupling $\lambda_f$. The green dashed line  indicates the estimated ballpark of $\gamma_\lambda$ in the MFPC scenario with two additional Dirac fermions in \labelcref{eq:estimategammalambda}. The green-shaded region shows a $20\%$ variation in the estimate of $h_{\cal B}$ in \labelcref{eq:hBapproxCW} and serves only as an indicative measure of the dependence of $\gamma_\lambda$ on $h_{\cal B}$.}
	\label{Fig:SMgammaNP}
\end{figure*}

Finally, the leading contribution in \labelcref{eq:gammalambdasimplified} remains to be determined. 
The momentum scale variation of the wave-function renormalisation can be obtained from the composite's two-point function 
\begin{align}
\Gamma^{\left(2\right) }_{{\cal B}\bar{\cal B},k}= \frac{\delta \Gamma_k}{\delta {\cal B}(p)\delta {\bar{ \cal B}}(-p)}&=  \imag Z_{{\cal B}}\, \sigma_\mu\, p_\mu \,.
\end{align}
Choosing the adequate projection onto the kinetic part of the two-point function and evaluating at vanishing external momenta, the anomalous dimension reads
\begin{align}\label{eq:etaB}
\gamma_{\cal B}=\frac{\partial_t  Z_{\cal B}}{  Z_{\cal B}}&= \left.  \frac{-\imag  \, \, \partial_{p^2}\left(\, \sigma_\mu\, p_\mu\,\, \partial_t \Gamma^{\left(2\right) }_{{\cal B}\bar{\cal B},k} \right)}{  Z_{\cal B}\,\tr \, [\sigma_\mu \sigma_\nu ]}\right|_{p=0} \,. 
\end{align}
The diagrammatic form of $\partial_t \Gamma^{(2)}_{{\cal B}\bar{\cal B},k}$ is depicted in the second row of \Cref{fig:twopointB}. The explicit expressions for the flow are provided in \Cref{app:flow}. The diagram with the fermionic flow, (A) in \Cref{fig:twopointB}, vanishes in the $p=0$ limit. To scrutinize this further we re-evaluated the diagram employing numerical regulators which implement a richer momentum shell integration. 
In fact, regulators featuring a more elaborated momentum shell structure always show diagram (A) to have a negligible contribution. Moreover, the analytic result for the flat or Litim regulator is properly reproduced from our numerical regulators, which supports the neglection of the contribution of diagram (A). 

Consequently, the anomalous dimension at vanishing external momentum is given by diagram (B) and reads (see \Cref{app:flow} for details)
\begin{align}\label{eq:gammaB}
\gamma_{{\cal B}}=& \,-\frac{ h^2_{{\cal B}}}{16\pi^2}\frac{N_{\rm TC}}{2} \tr[\mathcal{T}^2]\left(1+\frac{\gamma_{{\cal S}}}{5}\right)\,,
\end{align}
where $N_{\rm TC}$ is the dimension of the fundamental representation of the SU($N_{\rm TC})$-gauge group and $\gamma_{\cal S}$ is the anomalous dimension of the fundamental scalar TC-field coming from the cutoff derivative $\partial_t R_k$. The presence of the scalar anomalous dimension reflects the higher loop character of the fRG. 

Moreover, all parameters and couplings in \labelcref{eq:gammaB} are positive constants, leading to 
\begin{align}
\gamma_{\cal B}<0\,
\label{eq:boundgammaB}
\end{align}
as a natural limit. \Cref{eq:boundgammaB} reproduces the unitarity bound on fermionic fields \cite{Contino:2004vy} purely derived from the effective action. As discussed in \Cref{app:flow}, the choice of the regulator does not alter the sign of the anomalous dimension as this can be written as a total scale derivative of the one-loop diagram. See \labelcref{eq:signgammaB} for more details. However, $\gamma_{\cal B}$ could turn positive if the scalar anomalous dimension is largely negative, $\gamma_{\cal S} <-5$. This would require an extremely strong coupling leading to contributions from higher n-point functions becoming relevant. As this scenario is outside the scope of this computation, we consider the 1-loop expression and neglect the suppressed scalar anomalous dimension.

Collecting \labelcref{eq:gammafto0,eq:gammaqB0,eq:gammaB}, we find that the scaling of the linear mixing couplings is mainly determined by the anomalous dimension of the composite field 
\begin{align}\label{eq:gammalambdagammaB}
    \gamma_{\lambda}\approx -1 - \gamma_{\cal B}/2 \,,
\end{align}
leading to the relation 
\begin{align}\label{eq:msimplified}
m^f = &\frac{\mathcal{Y}_f \,v}{\sqrt{2}}\left( \frac{\Lambda_{\textrm{c}}}{\Lambda_{\textrm{UV}}}\right)^{  2 \gamma_\lambda} \sim \frac{\mathcal{Y}_f \,v}{\sqrt{2}}\left( \frac{\Lambda_{\textrm{c}}}{\Lambda_{\textrm{UV}}}\right)^{  -2 -\gamma_{\cal B}}\,,
\end{align} 
for the SM fermion masses. In \Cref{Fig:SMgammaNP} we have used~\labelcref{eq:msimplified} to illustrate the SM fermion masses as a function of the anomalous dimension $\gamma_\lambda$. Note that the parameter scan presented here is independent of the derivation of $\gamma_\lambda$, but subject to the phenomenological characteristics of the theory such as the size of the walking regime $\Delta\Lambda=\left(\Lambda_{\textrm{c}}/\Lambda_{\textrm{UV}}\right)$ and the value of fundamental couplings at the onset scale $(\Lambda_{\textrm{UV}})$. We will provide the connection to our results for the anomalous dimension further below. Nine different walking regimes of sizes $\log(\Delta\Lambda )\subset[-2,-10]$ are shown as clouds of points of different colours whose steepness increases with the regime's size. We consider the couplings $\overline{\lambda}_f$ and $g_{\cal B}$ to have a \textit{natural} value at the UV scale. Therefore, varying ${\cal Y}_f\subset [0.1,4 \pi]$, as defined in \labelcref{eq:Ynew}, covers the expected region. We note that the wider the range of ${\cal Y}_f$, the broader the spread of points and that ${\cal Y}_f \to 0$ for $y^{f}_{\rm TC} \to 0$, see \eqref{eq:ycy}.

For reproducing the SM flavour hierarchies within PC, different composites and scalings are necessary, as already reviewed in \Cref{sec:Introduction}. As the top Yukawa coupling is approximately unity, $\gamma_\lambda\sim 0$ is required, independently of the walking-regime size. This coupling could be predicted from the existence of the BZ fixed point, namely $\gamma_\lambda=0$. On the other hand, the lightest fermions require $\gamma_\lambda\sim 0.5-2.0$ which is subject to the size of the walking regime.  
As is apparent in \labelcref{eq:gammaB}, considering different excited radial, angular and spin states (encoded in ${\cal T}$) could facilitate generating the non-degeneracy. In this work, we do not discuss further the impact of these parameters and consider the particular case $\tr[\mathcal{T}^2]=1$.

In order to estimate the properties of the composite sector along the walking regime, the magnitude of the TC coupling, given by the fixed point solution of the beta function, is key. The MFPC matter content consists of 4 chiral fermions and 12 complex scalars transforming in the fundamental representation of SU(2)$_{\rm TC}$~\cite{Sannino:2016sfx,Cacciapaglia:2017cdi,Cacciapaglia:2020kgq}. Altogether, the TC-matter content affects the gauge beta function equivalently to the presence of 5 Dirac fermions. For such field content and representation, the SU(2)$_{\rm TC}$ gauge coupling is known to not show a BZ-like IR fixed point \cite{Dietrich:2006cm,Pica:2010xq}. 4-loop $\overline{\textrm{MS}}$ results show the BZ window to be present from 8 to 11 Dirac fermions. For fewer number of fermions (6 to 8), the IR FP exists but is not reached due to chiral symmetry breaking triggering before reaching the conformal regime~\cite{Appelquist:1996dq,Gies:2002hq,Braun:2006jd,Gies:2005as}. As both properties, conformal scaling and chiral symmetry breaking, are necessary for the current setup, we expect the desired dynamics in the boundary of the BZ window. Therefore we assume additional matter content charged under the TC-gauge group which decouples by the exit of the walking regime. Then, we consider the beta functions with 6, 7 and 8 Dirac fermions and employ the 4-loop $\overline{\textrm{MS}}$~\cite{vanRitbergen:1998pn,Vermaseren:2000nd,Dietrich:2006cm,Pica:2010xq,Fukano:2010yv} results to estimate $h_{\cal B}$. For it, we recall the discussion in \Cref{sec:Walkingregimes} and the estimate in \labelcref{eq:hBapproxCW}. Altogether, the estimated anomalous scaling with 1, 2 and 3 additional Dirac fermions reads
\begin{align}\label{eq:estimategammalambda}
    \gamma_\lambda&\sim -1 - \frac{1}{2}\left[-\frac{N_{\rm TC}}{32 \pi^2}\left(\frac{(1+\bar{m}_{\cal B}^2)\, r_g\, (g^*_{\rm TC})^4}{\bar{m}_{\cal B}}\right)\right]\,,\notag\\[1ex]
    &\sim \{4.74,\,0.47,\,-0.42 \}\,,
\end{align}
where we have assumed the scenario in which the gauge corrections are dominant over the Yukawa corrections, $g^*_{\rm TC}\gg y^*_{\rm TC}$ and considered $\bar{m}_{\cal B} =1$. Furthermore, we made the simplification $r_g\approx 1$. This rough estimate lacks a full determination of the flows of the couplings and combinatorial factors hence, must not be taken as conclusive for the validity of the models. Nevertheless, it serves at a conceptual level. 

In \Cref{Fig:SMgammaNP}, we depict the estimate for MFPC with two additional Dirac fermions with a green vertical dashed line. The green shaded region indicates a 20$\%$ variation in the estimate of $h_{\cal B}$ which serves only as a visual guide of the dependence of $\gamma_\lambda$ on $h_{\cal B}$.  

%%%%%%%%%%%%%%%%%%%%%%%%%%%%%%%%%%%
\section{Summary and conclusions}\label{sec:Conc}

In this work, we have explored the generation of the SM fermion mass hierarchies in the framework of FPC. Aside from a viable model building, two key features need to be satisfied for a realisation of the SM Yukawa couplings: the existence of a walking regime and an adequate scaling of the linear mixing couplings $\lambda_f$ between SM fermions and composites of the new sector. We have employed the non-perturbative fRG to address the latter point and determine whether the adequate scaling is realisable. 

In contrast to the original idea of PC, the bound states considered here are built out of a TC-fermion and a TC-scalar field, ${\cal B}\sim {\cal S} {\cal F}$. In \Cref{sec:EffAct,sec:EffActDynHad}, we introduced the effective action formalism particularised to FPC and derived in detail the hadronisation procedure for the composites of interest. This latter transformation consists in rewriting the two-TC-fermion two-TC-scalar 4-point function in a specific momentum channel as the exchange of the composite. To tackle the non-perturbative phenomenon of composite emergence and perform this field redefinition in a mathematically well-defined manner, we employ the fRG. This dynamical treatment of the emergence of resonances allows us to investigate the theory in a global and complete manner, taking into account fundamental and composite degrees of freedom simultaneously as well as their interplay. Accordingly, this formalism allows us to access the properties of the composites (eg. couplings, masses, ...) from the parameters of the fundamental effective action as shown in \Cref{sec:Walkingregimes}.

In \Cref{sec:FPCinEffAction} we derived the anomalous scaling of the linear mixing couplings from the momentum scaling of the 2-point functions in the effective action and resolved it in \Cref{sec:AnomalousDimCompositeSector} with the fRG. While the scaling of the vertex is found to vanish in the chiral limit, the composite anomalous dimension naturally reproduces the conformal bound. In the walking regime, we are able to reformulate the couplings of the composite action in terms of couplings of the fundamental FPC effective action. With this, we present an estimate for the MFPC scenario with two additional Dirac fermions coupled to the TC-gauge group. Lacking a full computation, this estimate does not allow us to conclusively judge the viability of the different models but furnishes a conceptual understanding of the underlying dynamics.

In general, we provide a novel application of functional methods to new physics scenarios involving strong dynamics and the emergence of composites. Using the fRG allows us to account for all quantum fluctuations and to investigate theories in a complete manner, taking into account fundamental and composite degrees of freedom simultaneously as well as their interplay. Importantly, the framework presented here provides the versatility needed for large theory space investigations required in model building. 

%%%%%%%%%%%%%%%%%%%%%%%%%%%%%%%%%%
\section*{Acknowledgements}

We thank A.~Angelescu, A.~Bally, G.~Cacciapaglia, Y.~Chung, F.~Sannino and M.~Yamada for discussions. This work is funded by the Deutsche Forschungsgemeinschaft (DFG, German Research Foundation) under Germany's Excellence Strategy EXC 2181/1 - 390900948 (the Heidelberg STRUCTURES Excellence Cluster) and under the Collaborative Research Centre SFB 1225 (ISOQUANT). 

\appendix 

%%%%%%%%%%%%%%%%%%%%%%%%%%%%%%%%%%%%%%
\section{Anomalous dimensions}
\label{app:flow}

In this Appendix, we provide the explicit computation of the composite two-point function flow equation, necessary for the determination of the anomalous dimension. In this computation we consider the TC fields to be massive to exemplify the mass-dependent nature of the fRG.

The necessary ingredients for the resolution of the diagrammatic flows in the second line of \Cref{fig:twopointB} are the Hessians or two-point point functions, and the full three-point functions. The former allows us access to the propagators of the fields in the theory. The regularised propagators for the dispersive modes of the different fields read,
\begin{align}\label{eq:propagators}
G_{\psi\bar{\psi},k}(p) &= \left[\Gamma^{\left(2\right) }_{\psi\bar{\psi},k}(p) +R^{{\psi}}_k (p)\right]^{-1}\notag\\
&= \frac{-\imag \sigma_\mu p_\mu \left(1+r^{(\psi)}\left(\frac{p^2}{k^2}\right)\right)}{Z_{\psi}\,\left( p^2\left(1+r^{(\phi)}\left(\frac{p^2}{k^2}\right)\right) +  m^2_{\psi}\right)}\,,
\end{align}
and
\begin{align}
G_{{\varphi}{\varphi}^\dagger,k} (p)&= \left[\Gamma^{(2)}_{{\varphi}{\varphi}^\dagger,k} +R^{\varphi)}_k \right]^{-1} \notag \\[1ex]
&= \left[Z_{\varphi}\,\left( p^2 \left(1+r^{(\phi)}\left(\frac{p^2}{k^2}\right)\right) + m^2_{\varphi}\right)\right]^{-1}\,.
\end{align}
where $\varphi$ stands for bosonic fields and $\psi$ for fermionic fields in the super field $\phi$. For the regulator functions $R_k$ we employ the \textit{flat} or \textit{Litim regulator} \cite{Litim:2000ci}
\begin{align}
	R^{\varphi}_k(p^2) =& Z_{\varphi}\, p^2 \,  r^{(\varphi)} (p^2/k^2)\,,\notag\\
    R^{\psi}_k(p^2) =& \imag\,  Z_{\psi}\,  \sigma_\mu p_\mu \,  r^{(\psi)} (p^2/k^2)\,,
\label{eq:regdeff}
\end{align}
with
\begin{align}\nonumber 
r^{(\varphi)} (x) =&  (-1+1/x)\,\theta\left(1-x\right)\,,\\[1ex]
r^{(\psi)} (x) =&  (-1+1/\sqrt{x})\,\theta\left(1-x\right)\,.
\label{eq:regxdeff}\end{align}
This regulator choice allows for analytical access to the flow of $n$-point functions at vanishing momentum. For fully quantitative studies smooth regulators (in momentum space) are better suited.

The full effective vertex can be read off from the effective action \labelcref{eq:Baction} by performing three functional derivatives,
\begin{align}
\Gamma^{(3)}_{{\cal S}{\cal F}\bar{\cal B},k}= \Gamma^{(3)}_{{\cal S}^\dagger{\cal B}\bar{{\cal F}},k}= h_{{\cal B}}\sqrt{Z_{{\cal B}}Z_{{\cal S}}Z_{{\cal F}}}\,.
\end{align}
The anomalous dimension can be obtained from the composite two-point function 
\begin{align}
 \partial_t \Gamma^{\left(2\right) }_{{\cal B}\bar{\cal B},k}(p) &= \partial_t \left[ Z_{{\cal B}}\, (\imag \sigma_\mu p_\mu + m_{\cal B})\right] \,,
\end{align}
leading to
\begin{align}
\gamma_{\cal B}=\frac{\partial_t  Z_{\cal B}}{  Z_{\cal B}}&= \left.  \frac{-\imag  \, \, \partial_{p^2}\left(\tr\, \Bigl[ \sigma_\mu p_\mu\, \partial_t \Gamma^{\left(2\right) }_{{\cal B}\bar{\cal B},k}(p)\Bigr] \right)}{  Z_{\cal B}\, \tr \, [\sigma_\mu \sigma_\nu ] }\right|_{p=0} \,. 
\end{align}
Collecting all pieces and integrating the loop momenta
\begin{widetext}
\begin{align}
    \gamma_{\cal B} &=-\frac{N_{\rm TC}}{2}
    h_{{\cal B}}^2\, \tr\left[\mathcal{T}^2\right] \Bigg\{\notag\\[2ex]
    &\hspace{1cm}+\partial_{p^2}\left[\int \frac{d^4 q}{(2\pi)^4}\frac{q^2 \left( \gamma_{{\cal F}} \left(k/q+1\right)-k/q \right)\theta \left(1-q^2/k^2\right)}{\left( (p+q)^2 (1+r^{(\phi)}((p+q)^2/k^2)) + m_{{\cal S}}^2\right)}\frac{(p\cdot q)(1+r^{(\psi)}(q^2/k^2))^2}{ \left(q^2(1+r^{(\psi)}(q^2/k^2))+ m_{{\cal F}}^2\right)^2 }\right]_{p=0}\notag \\[2ex]
    &\hspace{1cm}+ \partial_{p^2}\left[\int\frac{d^4 q}{(2\pi)^4}\frac{q^2 \left( \gamma_{{\cal S}} \left(k^2/q^2-1\right)+2k^2/q^2\right)\theta \left(1-q^2/k^2\right)}{\left( q^2 (1+r^{(\phi)}(q^2/k^2)) + m_{{\cal S}}^2\right)}\frac{(p\cdot(p+q)(1+r^{(\psi)}((p+q)^2/k^2))^2}{ \left((p+q)^2(1+r^{(\psi)}((p+q)^2/k^2))+ m_{{\cal F}}^2\right)^2 }\right]_{p=0}\Bigg\}\notag \\[2ex]
    &=-\frac{N_{\rm TC}}{2} h_{{\cal B}}^2\, \tr\left[\mathcal{T}^2\right] \Bigg\{\notag\\[2ex]   
   &\hspace{1cm}+\int \frac{d^4 q}{(2\pi)^4}\frac{k^2 q^2\,x_1^2 \,\left( \gamma_{{\cal F}} \left(k/q+1\right)-k/q\right)\theta(1-q^2/k^2)}{(q^2 +(k-q)(k+q)+ m_{{\cal F}}^2)^2(q^2 +(k-q)(k+q)+ m_{{\cal S}}^2)^2}\notag\\[2ex]
   &\hspace{10cm}\times\left[(1-q^2/k^2)\delta\left(-1+q^2/k^2\right)+(-1+\theta(1-q^2/k^2)\right] \notag \\[2ex]
   &\hspace{1cm}+\int \frac{d^4 q}{(2\pi)^4}\frac{ k q\,(-1+x_1^2)\, (k^2+ m_{{\cal F}}^2) \left( \gamma_{{\cal S}} \left(k^2/q^2-1\right)+2k^2/q^2\right) \theta(1-q^2/k^2)}{(q^2 +(k-q)(k+q)+ m_{{\cal F}}^2)^2(q^2 +(k-q)(k+q)+ m_{{\cal S}}^2)^2}\Bigg\}\notag \\[2ex]
    &=-\frac{ h^2_{{\cal B}}}{16\pi^2}\frac{N_{\rm TC}}{2} \tr[\mathcal{T}^2]\frac{(1+\gamma_{{\cal S}}/5)}{  \left(1+ \bar m_{{\cal F}}^2\right) \left(1+ \bar m_{{\cal S}}^2\right)^2}\,,
\label{eq:analyticflow}\end{align}
\end{widetext}
where the first term of each derivation step corresponds to diagram (A) and the second to (B) in \Cref{fig:twopointB}. $\gamma_{\cal S}$ and $\gamma_{\cal F}$ are the anomalous dimensions of the TC-scalar and TC-fermion fields respectively. In contrast with \labelcref{eq:gammaB}, the explicit contribution of the TC-field masses is here shown to illustrate the mass-dependent nature of the fRG. The integral $d^4 q$ has been performed in spherical coordinates 
\begin{align}
 d^4 q = dq \, dx_1 \, dx_2 \,d\phi\, \sqrt{1+x_1} q^3\,, 
\end{align}
with $x_i= \cos \theta_i$ integrated from $x_i \subset [-1,1]$. $\theta_1$ is the angle between the external $p$ and internal $q$ momenta in spherical coordinates and in the 2-point function, there is no dependence on a second external momenta $x_2$. Last $\phi$ is the azimuthal angle $\phi \subset [0,2\pi]$.

The first diagram (A), in the first line of \labelcref{eq:analyticflow} is always vanishing due to the momentum structure of the diagram. This can be read off from the first term in the second step of the derivation. This result is in agreement with previous computations of fermion-scalar systems \cite{Meibohm:2015twa,Pastor-Gutierrez:2022nki}.

One may employ other types of regulators which better implement the momentum shell integration. This computation was repeated for \textit{smooth}- and \textit{exponential}-type regulators. For all types of numerical regulators, we found the analytical result with \labelcref{eq:regdeff} in the respective limit. In the cases in which diagram (A) did not vanish, its contribution was negligible in comparison to diagram (B).

Aside from the regulator discussion, the result \labelcref{eq:boundgammaB} can be shown to remain unaffected by regulator choices. The two-point function flow (second line in \Cref{fig:twopointB}) can be rewritten as a total scale derivative of the uncut one-loop diagram. Consequently, the sign of the anomalous dimension is given by this diagram's contribution
\begin{align}\label{eq:signgammaB}
    \text{sign}\left[\gamma_{{\cal B}}\right]=\text{sign}\,   \left[\partial_t\parbox{2.cm}{\centering\begin{tikzpicture}[scale=0.35]\input{./figures/BBar_noReg.tex}\end{tikzpicture}}\right]<0
\end{align}
and is unchanged by a cut-off insertion. 

   \bibliography{references}
\end{document}

%% file: figures/4TCFermi.tex
\node (A1) at (0, 1) {};
\node (A2) at (0, -1) {};
\node (B1) at (1, 0) {};
\node (C1) at (2, 1) {};
\node (C2) at (2, -1) {};

\draw[fermion] (A1) -- (B1);
\draw[scalar] (B1) -- (A2);
\draw[fermion] (B1) -- (C1);
\draw[scalar] (C2) -- (B1);

\filldraw (1,0) circle (.15);

%% file: figures/4PFgluons.tex
\node (A1) at (0, 0.8) {};
\node (A2) at (0, -0.8) {};
\node (B1) at (1, 0.8) {};
\node (B2) at (2, 0.8) {};
\node (B11) at (1, -0.8) {};
\node (B22) at (2, -0.8) {};
\node (C1) at (3, 0.8) {};
\node (C2) at (3, -0.8) {};

 \draw[fermion] (A1) -- (B1) -- (B2) -- (C1); 
 \draw[scalar] (A2) -- (B11) -- (B22) -- (C2);
 \draw[gluon] (B1)--(B11);
 \draw[gluon] (B2)--(B22);

   \filldraw (B1) circle (.1); \filldraw (B2) circle (.1);
   \filldraw (B11) circle (.1); \filldraw (B22) circle (.1);

%% file: figures/4PFtriangle.tex
\node (A1) at (0, 0.8) {};
\node (A2) at (0, -0.8) {};
\node (B1) at (1, 0.8) {};
\node (B2) at (2, 0.8) {};
\node (B11) at (1.5, -0.8) {};
\node (B22) at (1.5, -0.8) {};
\node (C1) at (3, 0.8) {};
\node (C2) at (3, -0.8) {};

 \draw[fermion] (A1) -- (B1) -- (B2) -- (C1); 
 \draw[scalar] (A2) -- (B11) -- (B22) -- (C2);
 \draw[gluon] (B1)--(B11);
 \draw[gluon] (B2)--(B22);

   \filldraw (B1) circle (.1); \filldraw (B2) circle (.1);
   \filldraw (B11) circle (.1); \filldraw (B22) circle (.1);

%% file: figures/4PFSM.tex
\node (A1) at (0, 0.8) {};
\node (A2) at (0, -0.8) {};
\node (B1) at (1, 0.8) {};
\node (B2) at (2, 0.8) {};
\node (B11) at (1, -0.8) {};
\node (B22) at (2, -0.8) {};
\node (C1) at (3, 0.8) {};
\node (C2) at (3, -0.8) {};

 \draw[fermion] (A1) -- (B1);
 \draw[scalar] (B1) -- (B2);
 \draw[fermion] (B2) -- (C1); 
 \draw[scalar] (A2) -- (B11);
 \draw[fermion] (B11) -- (B22);
 \draw[scalar] (B22) -- (C2);
 \draw[fermionSM] (B1)--(B11);
 \draw[fermionSM] (B22)--(B2);

   \filldraw (B1)[draw=blue,fill=blue] circle (.1); \filldraw (B2)[draw=blue,fill=blue] circle (.1);
   \filldraw (B11)[draw=blue,fill=blue] circle (.1); \filldraw (B22)[draw=blue,fill=blue] circle (.1);

%% file: figures/4PFcomp.tex
  \node (A1) at (0, 1) {};
    \node (A2) at (0, -1) {};
    \node (B1) at (1, 0) {};
    \node (B2) at (2, 0) {};
    \node (C1) at (3, 1) {};
    \node (C2) at (3, -1) {};
 
    \draw[fermion] (A1) -- (B1);
    \draw[scalar] (A2) -- (B1);
    \draw[comp] (B1) -- (B2);
    \draw[fermion] (B2) -- (C1);
    \draw[scalar] (B2) -- (C2);
 
   \filldraw (1,0) circle (.12);
   \filldraw (2,0) circle (.12);
 % removes some leftover artifacts in the middle

%% file: figures/4TCFermiRest.tex
\node (A1) at (0, 1) {};
\node (A2) at (0, -1) {};
\node (B1) at (1, 0) {};
\node (C1) at (2, 1) {};
\node (C2) at (2, -1) {};

\draw[fermion] (A1) -- (B1);
\draw[scalar] (B1) -- (A2);
\draw[fermion] (B1) -- (C1);
\draw[scalar] (C2) -- (B1);

\filldraw (0.85,-0.15) rectangle ++(8pt,8pt);

%% file: figures/BPsiBar_fermionreg.tex
 % sunrise
	% loop:
	\draw [scalar] (1,0) arc (0:-180:1);
	\draw [fermion] (-1,0) arc (180:0:1);
%
	% regulator
	\filldraw [white] (0,1) circle (0.30);
	\draw [thick] (0,1) circle (0.30);
	\draw [thick] (0,1) -- ++(xyz polar cs:angle=45, radius=0.30);
	\draw [thick] (0,1) -- ++(xyz polar cs:angle=-135, radius=0.30);
	\draw [thick] (0,1) -- ++(xyz polar cs:angle=135, radius=0.30);
	\draw [thick] (0,1) -- ++(xyz polar cs:angle=-45, radius=0.30);
%
	% external legs
	\draw [comp] (-2.5,0) -- ++(xyz polar cs:angle=0, radius=1.5);
	
	\draw [fermionSM] (1,0) -- ++(xyz polar cs:angle=0, radius=1.5);
%
	% vertices
	\filldraw (-1,0) circle (.15)  ;
	\filldraw (1,0)[draw=blue,fill=blue] circle (.15) node[below=17pt,right=-25 pt]{ } node[below=-17pt,right=-25 pt]{ }  ;

%% file: figures/BPsiBar_scalarreg.tex
 % sunrise
	% loop:
	\draw [fermion] (-1,0) arc (0:180:-1);
	\draw [scalar] (-1,0) arc (180:0:1);
%
	% regulator
	\filldraw [white] (0,1) circle (0.30);
	\draw [thick] (0,1) circle (0.30);
	\draw [thick] (0,1) -- ++(xyz polar cs:angle=45, radius=0.30);
	\draw [thick] (0,1) -- ++(xyz polar cs:angle=-135, radius=0.30);
	\draw [thick] (0,1) -- ++(xyz polar cs:angle=135, radius=0.30);
	\draw [thick] (0,1) -- ++(xyz polar cs:angle=-45, radius=0.30);
%
	% external legs
	\draw [comp] (-2.5,0) -- ++(xyz polar cs:angle=0, radius=1.5);

	\draw [fermionSM] (1,0) -- ++(xyz polar cs:angle=0, radius=1.5);
%
	% vertices
	\filldraw (-1,0) circle (.15)  ;
	\filldraw (1,0)[draw=blue,fill=blue] circle (.15) node[below=17pt,right=-25 pt]{ } node[below=-17pt,right=-25 pt]{ }  ;

%% file: figures/BBar_fermionreg.tex
 % sunrise
	% loop:
	\draw [scalar] (1,0) arc (0:-180:1);
	\draw [fermion] (-1,0) arc (180:0:1);
%
	% regulator
	\filldraw [white] (0,1) circle (0.30);
	\draw [thick] (0,1) circle (0.30);
	\draw [thick] (0,1) -- ++(xyz polar cs:angle=45, radius=0.30);
	\draw [thick] (0,1) -- ++(xyz polar cs:angle=-135, radius=0.30);
	\draw [thick] (0,1) -- ++(xyz polar cs:angle=135, radius=0.30);
	\draw [thick] (0,1) -- ++(xyz polar cs:angle=-45, radius=0.30);
%
	% external legs
	\draw [comp] (-2.5,0) -- ++(xyz polar cs:angle=0, radius=1.5);
	
	\draw [comp] (1,0) -- ++(xyz polar cs:angle=0, radius=1.5);
%
	% vertices
	\filldraw (-1,0) circle (.15)  ;
	\filldraw (1,0) circle (.15) node[below=17pt,right=-25 pt]{ } node[below=-17pt,right=-25 pt]{ }  ;

%% file: figures/BBar_scalarreg.tex
 % sunrise
	% loop:
	\draw [fermion] (-1,0) arc (0:180:-1);
	\draw [scalar] (-1,0) arc (180:0:1);
%
	% regulator
	\filldraw [white] (0,1) circle (0.30);
	\draw [thick] (0,1) circle (0.30);
	\draw [thick] (0,1) -- ++(xyz polar cs:angle=45, radius=0.30);
	\draw [thick] (0,1) -- ++(xyz polar cs:angle=-135, radius=0.30);
	\draw [thick] (0,1) -- ++(xyz polar cs:angle=135, radius=0.30);
	\draw [thick] (0,1) -- ++(xyz polar cs:angle=-45, radius=0.30);
%
	% external legs
	\draw [comp] (-2.5,0) -- ++(xyz polar cs:angle=0, radius=1.5);

	\draw [comp] (1,0) -- ++(xyz polar cs:angle=0, radius=1.5);
%
	% vertices
	\filldraw (-1,0) circle (.15)  ;
	\filldraw (1,0) circle (.15) node[below=17pt,right=-25 pt]{ } node[below=-17pt,right=-25 pt]{ }  ;

%% file: figures/BBar_noReg.tex
 % sunrise
	% loop:
	\draw [scalar] (1,0) arc (0:-180:1);
	\draw [fermion] (-1,0) arc (180:0:1);
%
%
	% external legs
	\draw [comp] (-2.5,0) -- ++(xyz polar cs:angle=0, radius=1.5);
	
	\draw [comp] (1,0) -- ++(xyz polar cs:angle=0, radius=1.5);
%
	% vertices
	\filldraw (-1,0) circle (.25)  ;
	\filldraw (1,0) circle (.25) node[below=17pt,right=-25 pt]{ } node[below=-17pt,right=-25 pt]{ }  ;